\documentclass[twocolumn]{aastex62}

\usepackage{bm}
\usepackage{textcomp}
\usepackage{graphicx}
\usepackage{amssymb}
\usepackage{gensymb}
\usepackage{natbib}
\received{}
\revised{}
\accepted{}
\submitjournal{ApJ}

\shorttitle{Polarized Line Formation in Arbitrary Strength Magnetic Fields}
\shortauthors{Sampoorna et al.}
\begin{document}

\title{Polarized Line Formation in Arbitrary Strength Magnetic Fields\,:\\
{\textit{the case of a two-level atom with hyperfine structure splitting}}}
\correspondingauthor{M. Sampoorna}
\email{sampoorna@iiap.res.in}
\author{M. Sampoorna}
\affil{Indian Institute of Astrophysics, Koramangala,
Bengaluru 560 034, India}
\author{K. N. Nagendra}
\affiliation{Indian Institute of Astrophysics, Koramangala,
Bengaluru 560 034, India}
\affiliation{Istituto Ricerche Solari Locarno, Locarno Monti, Switzerland}
\author{K. Sowmya} 
\affiliation{Max-Planck-Institut f\"ur Sonnensystemforschung,
Justus-von-Liebig-Weg 3, 37077, G\"ottingen, Germany}
\author{J. O. Stenflo}
\affiliation{Istituto Ricerche Solari Locarno, Locarno Monti, Switzerland}
\affiliation{Institute for Particle Physics and Astrophysics, ETH Zurich, CH-8093 Zurich, Switzerland}
\author{L.~S. Anusha}
\affiliation{Max-Planck-Institut f\"ur Sonnensystemforschung,
Justus-von-Liebig-Weg 3, 37077, G\"ottingen, Germany}

\begin{abstract}
Quantum interference effects together with partial frequency redistribution 
(PFR) in line scattering produce subtle signatures in the so called 
Second Solar Spectrum (the linearly polarized spectrum of the Sun). 
These signatures are modified in the presence of arbitrary strength 
magnetic fields via the Hanle, Zeeman, and Paschen-Back effects. In the 
present paper we solve the problem of polarized line formation in a 
magnetized atmosphere taking into account scattering in a two-level atom 
with hyperfine structure splitting together with PFR. To this end we 
incorporate the collisionless PFR matrix derived in \citet{snss14} in the 
polarized transfer equation. We apply the scattering expansion method to solve 
this transfer equation. We study the combined effects of PFR and 
Paschen-Back effect on polarized line profiles formed in an isothermal 
one-dimensional planar atmosphere. For this purpose, 
we consider the cases of D$_2$ lines of Li\,{\sc i} and Na\,{\sc i}.  
\end{abstract}


\section{Introduction}
\label{pbhfsrt-intro}
The polarized spectra emanating from solar/ stellar atmospheres provide 
us with a unique diagnostic tool to unravel the underlying scattering 
physics and to detect the solar/ stellar magnetic fields. Therefore it is 
of utmost importance to solve the problem of polarized line formation in a 
magnetized atmosphere. In the present paper we consider this problem for the 
case of a two-level atom with unpolarized lower level and with hyperfine 
structure splitting (HFS). 

When a radiating atom possesses a non-zero nuclear spin ($I_s$), the fine 
structure states of the atom (designated by the total electronic angular 
momentum quantum number $J$) undergo hyperfine structure splitting. 
The $J-I_s$ coupling results in hyperfine structure states labeled by $F$ 
quantum number. The $F$ states belonging to a given $J$ state can interfere 
resulting in the so called $F$-state interference. The problem of polarized 
line formation including this $F$-state interference and partial frequency 
redistribution (PFR) in the non-magnetic case has been dealt with in 
\citet{flurietal03,holzreuteretal05,ssnss12,snss13,snsbr14,ssnrs13,lbjtb13}, 
and in \citet{btl15}. In the non-magnetic case, it is well-known that HFS 
causes a depolarization in the line core. In the presence of magnetic fields, 
the hyperfine structure $F$ states further split into magnetic substates. 
When the magnetic splitting is much smaller than the HFS, we are in the 
linear Hanle-Zeeman regime, wherein the magnetic splitting varies linearly 
with the field strength. When the magnetic splitting becomes comparable to or 
larger than the HFS, we enter the so-called incomplete Paschen-Back effect 
(PBE) regime. This field strength regime is characterized by non-linear 
splitting of the magnetic substates, the level-crossing, and anti-level 
crossing effects. With a further increase in the field strength, we enter the 
complete PBE regime, wherein the magnetic splitting again varies linearly 
with field strength. The different field strength regimes produce interesting 
signatures in the singly scattered polarized line profiles 
\citep[see][]{vb80,ll04,btl07,snss14,snss15}. Here we investigate the 
influence of PBE and PFR on Stokes profiles generated by multiple scattering 
in a magnetized atmosphere. 

A PFR matrix for scattering on a two-level atom with non-zero nuclear spin 
($I_s$) and in the presence of arbitrary magnetic fields has been derived 
for the collisionless case in \citet{lll97} using a metalevel formalism 
and in \citet{snss14} using a Kramers-Heisenberg scattering approach 
\citep{s94}. This PFR matrix takes into account Hanle, Zeeman, and PBE regimes 
of field strength. In the present paper we include this PFR matrix in the 
polarized transfer equation and present a numerical method of solution. 
The corresponding collisional PFR matrix is given in \citet{vb17}. Since 
the computation of the PFR matrix in the PBE regime is computationally 
expensive, we limit ourselves to the collisionless case. Furthermore, 
we consider the angle-averaged version of the magnetic PFR functions. 

Solving the problem of the transfer of polarized spectral line 
radiation in the presence of arbitrary 
magnetic fields and including PFR is computationally a complex task. 
\citet{sns08} generalized the perturbation method of \citet{nff02} to 
solve this problem for a two-level atom without HFS. More recently, 
an accelerated lambda iteration (ALI) method to solve the same problem 
is presented in \citet{abt17} and \citet{sns17}. An iterative method based on 
classical lambda iteration to solve the polarized PFR line transfer equation in 
arbitrary fields for a two-term atom without HFS is presented in 
\citet{dcm16}. In the present paper we apply 
the so-called scattering expansion method of \citet{fasn09} to solve the 
problem at hand. For our numerical studies we consider a one-dimensional 
isothermal constant property medium. We consider a two-level atom with upper 
level $J_b=3/2$ and lower level $J_a=1/2$ and with a nuclear spin $I_s=3/2$, 
which is representative of Na\,{\sc i} D$_2$ line. We also consider the 
Li\,{\sc i} D$_2$ line case (which also corresponds to $J=3/2 \to 1/2$ 
transition), wherein, $I_s=1$ for $^6$Li isotope and $I_s=3/2$ for $^7$Li 
isotope. For these two lines, we investigate the role played by 
the PBE, PFR, and radiative transfer. 

In Section~\ref{pbhfsrt-rte}, we present the governing equations of the 
problem at hand. The scattering expansion method is presented in 
Section~\ref{pbhfsrt-sem}. Numerical results are presented in 
Section~\ref{sec-results}. Conclusions are drawn in Section~\ref{sec-conclu}. 

\section{The Governing Equations}
\label{pbhfsrt-rte}
The polarized line radiative transfer equation in a planar atmosphere 
permeated by arbitrary magnetic fields ${\bm B}$ is given by 
\begin{equation}
\mu\frac{\partial}{\partial\tau}{\bm I}(\tau,x,{\bm n})=
{\bm K}
{\bm I}(\tau,x,{\bm n})-{\bm S}(\tau,x,{\bm n})\,. 
\label{pbhfsrt-e1}
\end{equation}
Here $d\tau=-k_L^{\rm A}dz$ is the line integrated vertical optical depth 
scale, $x=(\nu_{J_bJ_a}-\nu)/\Delta\nu_{\rm D}$ is the spectral 
distance from the line-center frequency $\nu_{J_bJ_a}$ (corresponding to a 
$J_b\to J_a$ transition in the absence of HFS and magnetic fields) in 
Doppler width ($\Delta\nu_{\rm D}$) units, ${\bm n} (\vartheta,\varphi)$ 
is the propagation direction of the ray (where $\vartheta$ is the inclination 
and $\varphi$ is the azimuth defined with respect to the atmospheric normal),
$\mu=\cos\vartheta$, and the Stokes vector ${\bm I}=[I,Q,U,V]^{\rm T}$. 
The frequency integrated line absorption coefficient is defined as 
\begin{equation}
k_L^{\rm A}={h\,\nu_{J_bJ_a} \over 4\pi}\,N_a\,B(J_aI_s \to J_bI_s),
\label{kla}
\end{equation}
where $N_a$ is the population of the lower level, $h$ is the Planck constant, 
and $B$ is the Einstein coefficient for absorption. 
The total absorption matrix is given by 
\begin{equation}
{\bm K} = {\bm \Phi} + r {\bm E}, 
\label{absorb-mat}
\end{equation}
where ${\bm E}$ is the $4\times4$ unity matrix and $r$ is 
the ratio of continuum to line averaged opacity. 
The $4\times4$ line absorption matrix ${\bm \Phi}$ is given by
\begin{equation}
{\bm \Phi}(x,{\bm n})=
\left(
\begin{array}{cccc}
\varphi_I & \varphi_Q & \varphi_U & \varphi_V\\
\varphi_Q & \varphi_I & \chi_V & -\chi_U\\
\varphi_U & -\chi_V & \varphi_I & \chi_Q\\
\varphi_V & \chi_U & -\chi_Q & \varphi_I\\
\end{array}
\right)\,.
\label{pbhfsrt-e2}
\end{equation}
For a two-level atom with HFS and with unpolarized lower level, the absorption 
coefficients $\varphi_i$ with $i=0,1,2,3$ (corresponding to $I$, $Q$, $U$, and 
$V$) and anomalous dispersion coefficients $\chi_i$ with $i=1,2,3$ 
(corresponding to $Q$, $U$, and $V$) can be found in \citet{ll04}. However 
these coefficients are usually given in a frame where the magnetic field is 
along the $Z$-axis. For the transfer computations, they need to be transformed 
to a frame where the $Z$-axis is along the atmospheric normal. Such a 
transformation is described in Appendix~B of \citet{sns17}. The resulting 
coefficients are then given by 
\begin{equation}
\varphi_i(x,{\bm n}) = \sum_{KQ} {\mathcal T}^{K}_{Q}(i,{\bm n})
{\rm e}^{-{\rm i}Q\varphi_B}
d^K_{Q0}(\vartheta_B)\Phi^{0K}_0(J_a,J_b,x),
\label{abs-coef-arf}
\end{equation}
\begin{equation}
\chi_i(x,{\bm n}) = \sum_{KQ} {\mathcal T}^{K}_{Q}(i,{\bm n})
{\rm e}^{-{\rm i}Q\varphi_B}
d^K_{Q0}(\vartheta_B)\Psi^{0K}_0(J_a,J_b,x),
\label{dis-coef-arf}
\end{equation}
where $\vartheta_B$ and $\varphi_B$ are the inclination and azimuth of the 
magnetic field with respect to the atmospheric normal. The symbol 
$d^K_{QQ'}(\vartheta_B)$ stands for reduced rotation matrices, which are 
tabulated in Table~2.1 of \citet{ll04}. $\mathcal{T}^K_Q(i,{\bm n})$ are the 
irreducible spherical tensors with $K=0,1,2$ and 
$-K\leqslant Q\leqslant +K$ \citep[see][]{landi84}. In analogy with the case 
of two-level atom without HFS we refer to $\Phi^{0K}_0$ and $\Psi^{0K}_0$ 
as the generalized profile function and generalized dispersion profile 
function respectively, but now they are defined for the case of a two-level 
atom with HFS. They are given by 
\begin{eqnarray}
&& \!\!\!\!\!\!\!\!
\Phi^{0K}_0(J_a,J_b,x) =
\frac{1}{2I_s+1}
\sqrt{3(2K+1)}\!\!\!\!\!
\sum_{i_ai_bF_aF_{a^\prime}F_bF_{b^\prime}}\nonumber \\ &&\!\!\!\!\!\!\!\!\times
\!\!\!\sum_{m_am_bq} (-1)^{q+1}
\sqrt{(2F_a+1)(2F_{a^\prime}+1)(2F_b+1)}\nonumber \\ &&\!\!\!\!\!\!\!\!\times
\sqrt{(2F_{b^\prime}+1)} C^{i_a}_{F_a}(J_aI_s,m_a)
C^{i_a}_{F_{a^\prime}}(J_aI_s,m_a)\nonumber \\ &&\!\!\!\!\!\!\!\!\times
C^{i_b}_{F_b}(J_bI_s,m_b)
C^{i_b}_{F_{b^\prime}}(J_bI_s,m_b) 
\left\lbrace
\begin{array}{ccc}
J_a & J_b & 1\\
F_b & F_a & I_s \\
\end{array}
\right\rbrace\nonumber \\ &&\!\!\!\!\!\!\!\!\times
\left\lbrace
\begin{array}{ccc}
J_a & J_b & 1\\
F_{b^\prime} & F_{a^\prime} & I_s \\
\end{array}
\right\rbrace
\left (
\begin{array}{ccc}
F_b & F_a & 1\\
-m_b & m_a & -q \\
\end{array}
\right )
\left (
\begin{array}{ccc}
F_{b^\prime} & F_{a^\prime} & 1\\
-m_b & m_a & -q \\
\end{array}
\right )\nonumber \\ &&\!\!\!\!\!\!\!\!\times 
\left (
\begin{array}{ccc}
1 & 1 & K\\
q & -q & 0 \\
\end{array}
\right ) 
H(a,x_{ba})\ ,
\label{gen-prof-hfs}
\end{eqnarray}
and
\begin{equation}
\Psi^{0K}_0(J_a,J_b,x)
=\Phi^{0K}_0(J_a,J_b,x)
\{{H(a,x_{ba}) \rightarrow F(a,x_{ba})}\}.
\label{gen-disp-prof-hfs}
\end{equation}
Here $m$ denotes the magnetic quantum number, and 
$\{{H(a,x_{ba}) \rightarrow F(a,x_{ba})}\}$ means that the $H(a,x_{ba})$ 
in Eq.~(\ref{gen-prof-hfs}) should be replaced by $F(a,x_{ba})$ in 
Eq.~(\ref{gen-disp-prof-hfs}). In the incomplete PBE regime, the magnetic 
field produces a mixing of the $F$-states belonging to a given $J$, so that 
$F$ is no longer a good quantum number. Thus the basis vectors in the 
incomplete PBE regime are denoted by $|JI_sim\rangle$, where the symbol $i$ 
labels the different states spanned by the quantum numbers ($J$, $I_s$, $m$). 
The symbol $C^i_F(JI_s,m)$ denotes the expansion coefficients corresponding 
to the expansion of the incomplete PBE regime basis vectors $|JI_sim\rangle$ 
on to the linear Hanle-Zeeman regime basis vectors $|JI_sFm\rangle$ 
\citep[see e.g., Eq.~(11) of][]{snss14}. They are obtained by 
diagonalizing the magnetic Hamiltonian in the incomplete PBE regime 
\citep[see][also \citealt{snss14}]{ll04}. We remark that, 
Eqs.~(\ref{gen-prof-hfs}) and (\ref{gen-disp-prof-hfs}) correspond to the 
observer's reference frame, wherein the Doppler motion of atoms through a 
convolution of the Maxwellian velocity distribution with the atomic frame 
profile function has been accounted for. Thus, in these equations, 
$H(a,x_{ba})$ and $F(a,x_{ba})$ denote respectively the magnetically shifted 
Voigt and Faraday-Voigt functions, wherein $a$ is the damping parameter and 
$x_{ba}=(\nu_{i_bm_b,i_am_a}
-\nu)/\Delta\nu_{\rm D}$ with $\Delta\nu_{\rm D}$ the Doppler width and 
$\nu_{i_bm_b,i_am_a}$ the line center frequency for $i_bm_b \to i_am_a$ 
transition and it is defined as
\begin{eqnarray}
&& \!\!\!\!\!\!\!\!\!\!\!\!\!\!\nu_{i_bm_b,i_am_a}=\nu_{J_bI_si_bm_b,J_aI_si_am_a}
\nonumber \\ &&
\!\!\!\!\!\!\!\!\!\!\!\!=\nu_{J_bJ_a} + \frac{E_{i_b}(J_bI_s,m_b)
-E_{i_a}(J_aI_s,m_a)}{h}\,.
\label{pbhfsrt-e4a}
\end{eqnarray}
Here $E$ is the energy shift of a magnetic substate measured with respect 
to the energy of the parent $J$ state \citep[see][for details on the way 
the energy shifts are calculated]{snss14}. 

The source vector ${\bm S}(\tau,x,{\bm n})$ appearing in 
Eq.~(\ref{pbhfsrt-e1}) is given by
\begin{equation}
{\bm S}(\tau,x,{\bm n})=
(r{\bm E}+\epsilon{\bm \Phi})B_{\nu_{J_bJ_a}}{\bm U}
+{\bm S}_{\rm scat}(\tau,x,{\bm n})\,.
\label{pbhfsrt-e5}
\end{equation}
Here, $B_{\nu_{J_bJ_a}}$ is the Planck function, ${\bm U}=[1,0,0,0]^{\rm T}$, 
and
\begin{eqnarray}
&&\!\!\!\!\!\!\!\!\!\!\!\!\!\!\!\!\!\!\!\!\!\!{\bm S}_{\rm scat}(\tau,x,{\bm n})=(1-\epsilon)\nonumber \\ &&
\!\!\!\!\!\!\!\!\!\!\!\!\!\!\!\!\!\!\!\!\!\times\oint \frac{d{\bm n^\prime}}{4\pi}
\int_{-\infty}^{+\infty}\!\!\!\!\!\! dx^\prime {\bm R}(x,{\bm n},x^\prime,
{\bm n^\prime},{\bm B}){\bm I}(\tau,x^\prime,{\bm n^\prime})\,,
\label{pbhfsrt-e6}
\end{eqnarray}
where $\epsilon$ is the thermalization parameter. In the absence of elastic 
collisions the redistribution matrix ${\bm R}$ is simply given by the type-II 
redistribution matrix ${\bm R}^{\rm II}$. For a two-level atom with HFS and 
in the PBE regime this matrix is given in Eq.~(16) of \citet{snss14}. However 
it is defined in a frame where the magnetic field is along the 
$Z$-axis. As described in Appendix~A of \citet{sns17}, we transform this matrix 
to the atmospheric reference frame (where $Z$ axis is along the atmospheric 
normal). The $ij$th element of this matrix then takes the form 
\begin{eqnarray}
&&\!\!\!\!\!\!\!
{R}^{\rm II}_{ij}(x,{\bm n}, x^\prime,{\bm n}^\prime,{\bm B})=
\sum_{KQ}{\mathcal T}^{K}_Q(i,{\bm n})
\sum_{K^\prime Q^\prime} \nonumber \\ &&\!\!\!\!\!\!\!\times
N^{K,K^\prime}_{QQ^\prime,{\rm II}}(x,x^\prime,\Theta,{\bm B}) 
(-1)^{Q^\prime}  
{\mathcal T}^{K^{\prime}}_{-Q^\prime}(j,{\bm n}^\prime), 
\label{r2-arf}
\end{eqnarray}
where ${\bm n} (\vartheta, \varphi)$ and ${\bm n}^\prime (\vartheta^\prime, 
\varphi^\prime)$ refer respectively to the scattered and incident ray 
directions with respect to the atmospheric normal, and $\Theta$ denotes the 
scattering angle between the incident and scattered rays. The type-II magnetic 
kernel has the form 
\begin{eqnarray}
&&\!\!\!\!\!\!\!\!\!\!\!\!\!\!\!\!
N^{K,K^\prime}_{QQ^\prime,{\rm II}}(x,x^\prime,\Theta,{\bm B}) 
= {\rm e}^{{\rm i}(Q^\prime-Q)\varphi_B} \sum_{Q^{\prime\prime}} 
d^{K}_{QQ^{\prime\prime}}(\vartheta_B) 
\nonumber \\ &&\!\!\!\!\!\!\!\!\!\!\!\!\!\!\!\!\times
{\mathcal R}^{K,K^\prime}_{Q^{\prime\prime},{\rm II}}(x,x^\prime,\Theta,B)
d^{K^{\prime}}_{Q^{\prime\prime}Q^\prime}(-\vartheta_B)\,.
\label{n2}
\end{eqnarray}
The PFR functions 
${\mathcal R}^{K,K^\prime}_{Q^{\prime\prime},{\rm II}}(x,x^\prime,\Theta,B)$
for the case of a two-level atom with HFS and in the PBE regime are given by 
\begin{eqnarray}
&&\!\!\!\!\!\!\!\!\!\!
{\mathcal R}^{K,K^\prime}_{Q^{\prime\prime},{\rm II}}(x,x^\prime,\Theta,B)=
\frac{3(2J_b+1)}{(2I_s+1)}
\sum_{i_am_ai_fm_fi_bm_bi_{b^\prime}m_{b^\prime}}
\nonumber \\ && \!\!\!\!\!\!\!\!\!\! \times
\sqrt{(2K+1)(2K^\prime+1)}
\cos\beta_{i_{b^\prime}m_{b^\prime}i_bm_b}
{\rm e}^{{\rm i}\beta_{i_{b^\prime}m_{b^\prime}i_bm_b}}
\nonumber \\ && \!\!\!\!\!\!\!\!\!\! \times
[(h^{\rm II}_{i_bm_b,i_{b^\prime}m_{b^\prime}})_{i_am_ai_fm_f}+
{\rm i}(f^{\rm II}_{i_bm_b,i_{b^\prime}m_{b^\prime}})_{i_am_ai_fm_f}]
\nonumber \\ && \!\!\!\!\!\!\!\!\!\! \times 
\sum_{F_aF_{a^\prime}F_fF_{f^\prime}F_bF_{b^\prime}
F_{b^{\prime\prime}}F_{b^{\prime\prime\prime}}}
\sum_{qq^\prime q^{\prime\prime}
q^{\prime\prime\prime}}
(-1)^{q-q^{\prime\prime\prime}+Q^{\prime\prime}}
\nonumber \\ && \!\!\!\!\!\!\!\!\!\! \times
\sqrt{(2F_a+1)(2F_f+1)(2F_{a^\prime}+1)(2F_{f^\prime}+1)}
\nonumber \\ && \!\!\!\!\!\!\!\!\!\! \times
\sqrt{(2F_b+1)(2F_{b^\prime}+1)(2F_{b^{\prime\prime}}+1)
(2F_{b^{\prime\prime\prime}}+1)}
\nonumber \\ && \!\!\!\!\!\!\!\!\!\! \times
C^{i_f}_{F_f}(J_aI_s,m_f) 
C^{i_f}_{F_{f^\prime}}(J_aI_s,m_f) 
C^{i_a}_{F_a}(J_aI_s,m_a)
\nonumber \\ && \!\!\!\!\!\!\!\!\!\! \times
C^{i_a}_{F_{a^\prime}}(J_aI_s,m_a)
C^{i_b}_{F_b}(J_bI_s,m_b) 
C^{i_b}_{F_{b^{\prime\prime}}}(J_bI_s,m_b)
\nonumber \\ && \!\!\!\!\!\!\!\!\!\! \times
C^{i_{b^\prime}}_{F_{b^\prime}}(J_bI_s,m_{b^\prime})
C^{i_{b^\prime}}_{F_{b^{\prime\prime\prime}}}(J_bI_s,m_{b^\prime})
\left\lbrace
\begin{array}{ccc}
J_a & J_b & 1\\
F_b & F_f & I_s \\
\end{array}
\right\rbrace
\nonumber \\ && \!\!\!\!\!\!\!\!\!\! \times
\left\lbrace
\begin{array}{ccc}
J_a & J_b & 1\\
F_{b^\prime} & F_{f^\prime} & I_s \\
\end{array}
\right\rbrace
\left\lbrace
\begin{array}{ccc}
J_a & J_b & 1\\
F_{b^{\prime\prime}} & F_a & I_s \\
\end{array}
\right\rbrace
\left\lbrace
\begin{array}{ccc}
J_a & J_b & 1\\
F_{b^{\prime\prime\prime}} & F_{a^\prime} & I_s \\
\end{array}
\right\rbrace
\nonumber \\ && \!\!\!\!\!\!\!\!\!\! \times
\left (
\begin{array}{ccc}
F_b & F_f & 1\\
-m_b & m_f & -q \\
\end{array}
\right )
\left (
\begin{array}{ccc}
F_{b^\prime} & F_{f^\prime} & 1\\
-m_{b^\prime} & m_f & -q^{\prime} \\
\end{array}
\right )
\left (
\begin{array}{ccc}
F_{b^{\prime\prime}} & F_a & 1\\
-m_b & m_a & -q^{\prime\prime} \\
\end{array}
\right )
\nonumber \\ && \!\!\!\!\!\!\!\!\!\! \times
\left (
\begin{array}{ccc}
F_{b^{\prime\prime\prime}} & F_{a^\prime} & 1\\
-m_{b^\prime} & m_a & -q^{\prime\prime\prime} \\
\end{array}
\right )
\left (
\begin{array}{ccc}
1 & 1 & K\\
q & -q^{\prime} & Q^{\prime\prime} \\
\end{array}
\right )
\left (
\begin{array}{ccc}
1 & 1 & K^\prime\\
q^{\prime\prime\prime} & -q^{\prime\prime} & -Q^{\prime\prime}\\
\end{array}
\right )\,.
\label{pbhfsrt-e7}
\end{eqnarray}
The auxiliary functions $h^{\rm II}$ and $f^{\rm II}$ are defined in 
Equations~(18)--(22) of \citet{snss14}. All the different symbols and 
quantities appearing in the above equation can be found in the same 
reference. We recall that Eq.~(\ref{pbhfsrt-e7}) is defined 
in the observer's reference frame, wherein the effects of Doppler 
redistribution have been account for.

From Eq.~(\ref{pbhfsrt-e1}) it is clear that all the four Stokes parameters 
are coupled to each other. We apply the DELOPAR method of 
\citet[][see also \citealt{sns08}]{jtb03} to obtain a formal solution of such 
a coupled first order ordinary differential equation. 

\section{Scattering Expansion Method}
\label{pbhfsrt-sem}
\citet{fasn09} presented an iterative method to solve the problem of 
polarized line formation in weak magnetic fields. This method is based 
on the Neumann series expansion of the components of the source vector 
that contribute to polarization. Such a series is equivalent to an 
expansion in the mean number of scattering events and hence the name scattering 
expansion method. This method was originally developed for transfer 
computations with complete frequency redistribution in \citet{fasn09} 
and has been 
extended to include (i) angle-dependent PFR in \citet{snf11} for the 
non-magnetic case and in \citet{ns11,ssnra13} for the weak field Hanle effect,
(ii) angle-averaged PFR in \citet{sns12} for the weak field Hanle effect, 
(iii) non-coherent electron scattering in \citet{snsr12}, and (iv) quantum 
interference with angle-dependent PFR in \citet{ssnrs13}. In the present 
section, we extend the scattering expansion method to the problem 
at hand, namely, polarized line formation in arbitrary magnetic fields. 

We write the source vector given in Equation~(\ref{pbhfsrt-e5}) in the 
component form as
\begin{eqnarray}
&&\!\!\!\!\!\!\!\!\!\! 
{S}_{i} (\tau,x,{\bm n})= 
\sum_{j=0}^3\bigg[(r{E}_{ij}+\epsilon{\Phi}_{ij})
B_{\nu_{J_bJ_a}}{U}_j +(1-\epsilon)
\nonumber \\ &&\!\!\!\!\!\!\!\!\!\! \times\oint \frac{d{\bm n^\prime}}{4\pi}
\int_{-\infty}^{+\infty} dx^\prime 
R_{ij}(x,{\bm n},x^\prime,{\bf n}^\prime,{\bm B})
I_j(\tau,x^\prime,{\bm n}^\prime)\bigg].
\label{pbhfsrt-e11}
\end{eqnarray}
In the scattering expansion method, we first compute the Stokes $I$ by 
neglecting its coupling to $Q$, $U$, and $V$. Thus the transfer 
equation~(\ref{pbhfsrt-e1}) for Stokes $I$ reduces to 
\begin{equation}
\mu\frac{\partial}{\partial\tau}{\tilde I}(\tau,x,{\bm n})=(\varphi_I+r)
{\tilde I}(\tau,x,{\bm n})-{\tilde S}_{I}(\tau,x,{\bm n})\,, 
\label{sem-rte-i}
\end{equation}
where tilde denotes approximate values. The approximate source 
function is given by
\begin{eqnarray}
&& \!\!\!\!\!\!\!\!\!\!
{\tilde S}_{I} (\tau,x,{\bm n})= 
(r+\epsilon\varphi_I)B_{\nu_{J_bJ_a}} \!\!
+(1-\epsilon)\nonumber \\ &&\!\!\!\!\!\!\!\!\!\!
\times\oint \frac{d{\bm n^\prime}}{4\pi}
\int_{-\infty}^{+\infty} dx^\prime 
R_{00}(x,{\bm n},x^\prime,{\bf n}^\prime,{\bm B})
\tilde I(\tau,x^\prime,{\bm n}^\prime). 
\label{pbhfsrt-e12}
\end{eqnarray}
It is interesting to note that the approximations made to 
obtain Eqs.~(\ref{sem-rte-i}) and (\ref{pbhfsrt-e12}) are similar to the 
polarization free approximation of \citet{tl96}. 
Equations~(\ref{sem-rte-i}) and (\ref{pbhfsrt-e12}) can now be solved using 
an ALI method similar to that described in \citet{sns17}. 
The ALI method is based on the introduction of an approximate 
lambda operator ${\bf \Lambda}^\ast_{x'{\bm n}'}$, which is chosen to be 
the diagonal of the true lambda operator \citep[see][]{oab86}. The ALI 
iterations are initiated by choosing ${\tilde S}_{I} (\tau,x,{\bm n})=
(r+\epsilon\varphi_I)B_{\nu_{J_bJ_a}}$. At each step in the iteration process, 
a formal solution of Eq.~(\ref{sem-rte-i}) using the short-characteristic 
technique of \citet{ok87} {\textbf{provides an updated}} estimate of 
${\tilde S}^{(n)}_{I}$, where the superscript $(n)$ refers to the $n$th 
iteration step. The source function corrections 
$\delta \tilde {\bm S}^{(n)}_I$ at each ALI iteration step are computed by 
solving a system of linear equations
\begin{equation}
{\bf A}\,\delta {\tilde {\bm S}}^{(n)}_I = {\bm r}^{(n)},
\label{deltas}
\end{equation}
where ${\bm r}^{(n)}$ is a residual vector 
\citep[given by the right-hand-side of Eq.~(18) of][]{sns17}. At each depth 
point, $\delta \tilde {\bm S}^{(n)}_I$ and ${\bm r}^{(n)}$ are vectors of
length $N_\lambda\,
2N_\mu\,N_{\varphi}$, where $N_\lambda$ is the number of wavelength points 
in the range $[\lambda_{\rm min},\lambda_{\rm max}]$, $N_\mu$ is the number 
of angle points in the range $[0<\mu\leqslant 1]$, and $N_{\varphi}$ is the 
number of azimuth points in the range $[0\leqslant\varphi\leqslant 2\pi]$. 
Since we use diagonal approximate lambda 
operator, the matrix ${\bf A}$ is also diagonal on the spatial grid. However, 
at each depth point, the matrix ${\bf A}$ has the dimension of $(N_\lambda\,
2N_\mu\,N_{\varphi}\,\times\,N_\lambda\,2N_\mu\,N_{\varphi})$. 
We remark that, when angle-averaged version of 
${\mathcal R}^{K,K^\prime}_{Q^{\prime\prime},{\rm II}}$ (see 
Eq.~(\ref{pbhfsrt-e7})) is used, it is possible to reduce the dimension of 
the matrix ${\bf A}$ to $(N_\lambda\,\times\,N_\lambda)$ at each depth point, 
following a technique described in \citet{abt17}. Basically 
Eq.~(\ref{pbhfsrt-e12}) can be expanded in terms of 
${\mathcal T}^{K}_{Q}(0,{\bm n})$ and reduced source functions that depend 
only on frequency or wavelength. However, since we continue to work with 
Eq.~(\ref{pbhfsrt-e12}) as it is, the size of the matrix ${\bf A}$ remains 
the same irrespective of whether we use angle-dependent or angle-averaged 
version of ${\mathcal R}^{K,K^\prime}_{Q^{\prime\prime},{\rm II}}$. On the 
other hand, to reduce the memory requirements we avoid storing the matrix 
${\bf A}$ on the main memory and use secondary storage. 

Clearly, solving Eq.~(\ref{deltas}) is the 
most time consuming part of the problem, apart from the computation of 
${\bm R}^{\rm II}$ matrix and ${\bm S}_{\rm scat}$ given in 
Eq.~(\ref{pbhfsrt-e6}). Traditionally, the system of linear equations 
represented by Eq.~(\ref{deltas}) is solved using a frequency-by-frequency 
(FBF) method of \citet{pa95}, which has been generalized in \citet{snf11} to 
handle angle-dependent PFR matrix. When the dimension of matrix ${\bf A}$ is 
small one can use one step matrix inversion techniques such as the 
LU decomposition scheme \citep[see e.g.,][]{pft86} to solve Eq.~(\ref{deltas}). 
However, when $N_\lambda$, $N_\mu$, and $N_{\varphi}$ are large (which is 
required for accurate computation of the PFR matrix in the PBE regime), the 
FBF method becomes computationally very expensive. Therefore, here we solve 
Eq.~(\ref{deltas}) by an iterative method in a Krylov subspace, called 
the restarted generalized minimum residual (mGMRES) method originally 
developed by \citet[][see also \citealt{hm14}]{ss86}. For the problem 
at hand, with $N_\lambda=121$, $N_\mu=7$, and $N_{\varphi}=8$ the mGMRES 
method is about 15 times faster than the FBF method. The ALI 
iterations are stopped when the maximum relative change on 
${\tilde S}_{I} (\tau,x,{\bm n})$ becomes smaller than $10^{-4}$.

Retaining only the contribution of $I$ on the {\textbf{right-hand-side}} of 
Eq.~(\ref{pbhfsrt-e11}) 
for the $i=1,2,$ and $3$ components of $S_{i}$, we obtain the single 
scattering approximation to these components as 
\begin{eqnarray}
&& \!\!\!\!\!\!\!\!\!\![{\tilde S}_{i} (\tau,x,{\bm n})]^{(1)}= 
\epsilon\varphi_i B_{\nu_{J_bJ_a}} 
+(1-\epsilon)\nonumber \\ && \!\!\!\!\!\!\!\!\!\!\times 
\oint \frac{d{\bm n^\prime}}{4\pi}
\int_{-\infty}^{+\infty} dx^\prime 
R_{i0}(x,{\bm n},x^\prime,{\bf n}^\prime,{\bm B})
\tilde I(\tau,x^\prime,{\bm n}^\prime),
\label{pbhfsrt-e13}
\end{eqnarray}
where $\tilde I$ is the solution obtained by solving 
Eqs.~(\ref{sem-rte-i}) and (\ref{pbhfsrt-e12}) by an ALI method described 
above. In Eq.~(\ref{pbhfsrt-e13}), 
$i=1,2,3$ and the superscript $(1)$ stands for single scattering. 
This single scattered Stokes source vector is used in 
a DELOPAR formal solver to obtain the Stokes parameters after the first 
scattering, namely $[\tilde I_i(\tau,x,{\bm n})]^{(1)}$. These Stokes parameter 
values are then used in Equation~(\ref{pbhfsrt-e11}) in an iterative 
sequence based on the orders of scattering approach to calculate the 
emergent Stokes parameters. Thus, the iterative sequence for the $k$th 
scattering is given by
\begin{eqnarray}
&& \!\!\!\!\!\!\!\!\!\!
[{\tilde S}_{i} (\tau,x,{\bm n})]^{(k)}=
[{\tilde S}_{i} (\tau,x,{\bm n})]^{(1)}
+(1-\epsilon)
\oint \frac{d{\bm n^\prime}}{4\pi} 
\nonumber \\ &&\!\!\!\!\!\!\!\!\!\!\times
\int_{-\infty}^{+\infty} \!\!\!\!\!dx^\prime 
\sum_{j=1}^3 R_{ij}(x,{\bm n},x^\prime,{\bf n}^\prime,{\bm B})
[\tilde I_j(\tau,x^\prime,{\bm n}^\prime)]^{(k-1)},
\label{pbhfsrt-e16}
\end{eqnarray}
where $i=1,2,3$. The iterations are stopped when the maximum relative 
change on the surface polarization becomes smaller than $10^{-4}$ 
\citep[see also][]{snf11}. 

\begin{deluxetable*}{ccccccccc}
\tablecolumns{8}
\tablewidth{0pt}
\tablecaption{Atomic parameters corresponding to the D$_2$ lines of Li\,{\sc i} 
and Na\,{\sc i}.
\label{atomic-parameters}}  
\vspace{5.mm}
\tablehead{
\colhead{Isotope} & \colhead{Abundance (\%)} & \colhead{$I_s$} & 
\colhead{$\lambda$ (\AA)} & \colhead{$A_{ba}\times 10^7$ (s$^{-1}$)} & 
\colhead{Isotope shift (MHz)} & \multicolumn{3}{c}{HFS constants (MHz)} \\
\cline{7-9}
\colhead{} & \colhead{} & \colhead{} &
\colhead{} & \colhead{} & \colhead{} &
\multicolumn{2}{c}{$\mathcal{A}_J$} & \colhead{$\mathcal{B}_J$}\\
\cline{7-8}
\colhead{} & \colhead{} & \colhead{} &
\colhead{} & \colhead{} & \colhead{} &
\colhead{$^2$S$_{1/2}$} & \colhead{$^2$P$_{3/2}$} & \colhead{$^2$P$_{3/2}$} \\
}
\startdata
$^6$Li & 7.59 & 1& 6707.90232 & 3.689 &$-$10534.93 & 152.136 & $-$1.155 
       & $-$0.010\\
$^7$Li & 92.41 & 3/2 & 6707.74416 & 3.689 & Reference & 401.752 & $-$3.055
       & $-$0.221\\
       &       &     &            &       & Isotope   &         &         
       &         \\
Na & 100. & 3/2 & 5889.95095 & 6.3 & - & 885.810 & 18.534
   & $+$2.724\\
\enddata
\end{deluxetable*}

\section{Numerical Results and Discussions}
\label{sec-results}

\subsection{Atomic and Atmospheric Models}
For the studies presented in this paper, we consider the atomic systems 
corresponding to the D$_2$ lines of Li\,{\sc i} and Na\,{\sc i}. 
The D$_2$ line results from a $J_b=3/2 \to J_a=1/2$ transition. The atomic 
parameters, such as the HFS ${\mathcal A}_J$ and ${\mathcal B}_J$ 
constants, Einstein $A_{ba}$ coefficients, and isotope shifts are taken from 
\citet{blt09} for the case of Li\,{\sc i} and from \citet{steck03} 
for the case of Na\,{\sc i}. 
For convenience, these parameters are given in Table~\ref{atomic-parameters}. 
{{\textbf{Note that $\mathcal{B}_J$ for $^2$S$_{1/2}$ lower-level is 
zero.}}}
\begin{figure*}[!ht]
\centering
\includegraphics[scale=0.8]{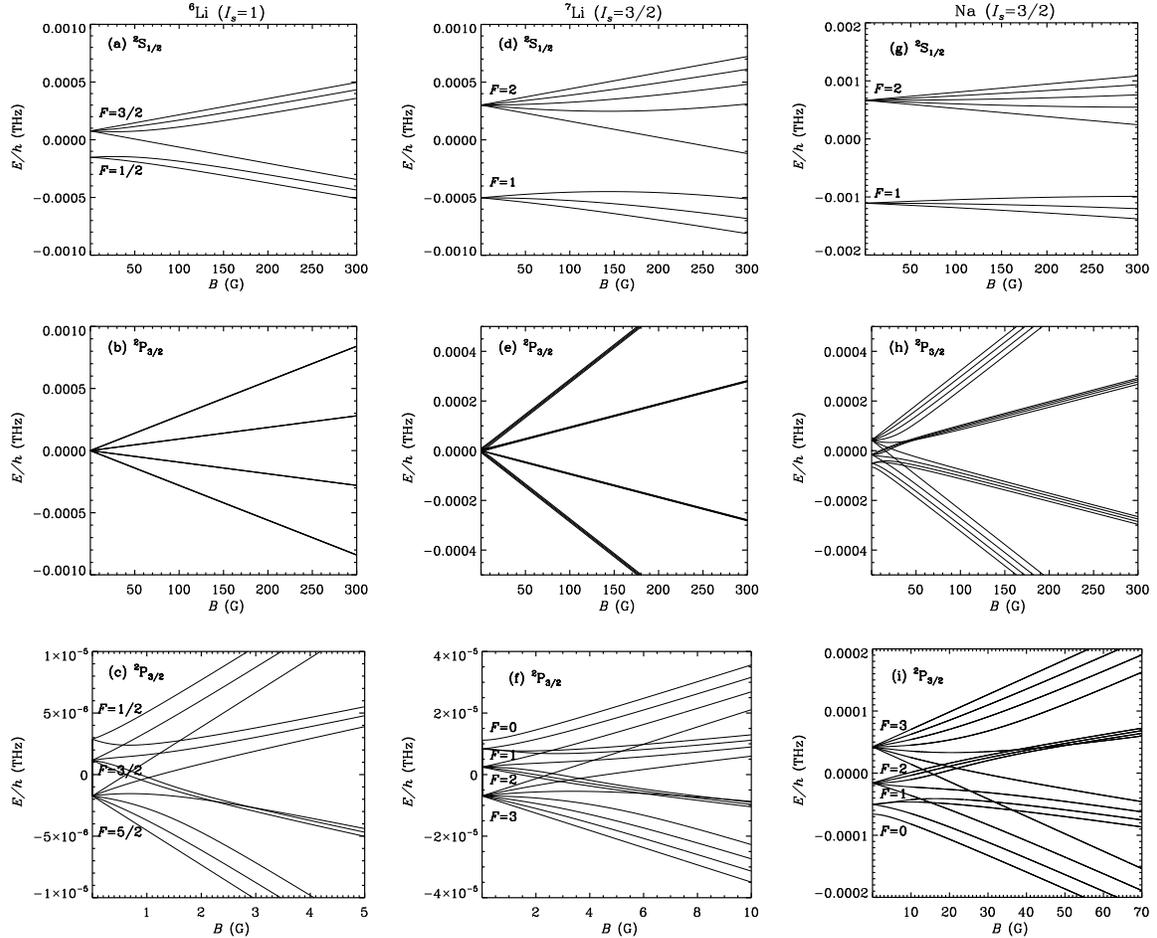}
\caption{Level-crossing diagrams corresponding to the lower (top panels) and 
upper (middle and bottom panels) levels of D$_2$ lines of $^6$Li (left column), 
$^7$Li (middle column), and Na (right column). The bottom panels represent a 
zoom-in of the middle panels to clearly show the various level-crossings in the 
incomplete PBE regime.}
\label{level-cross-fig}
\end{figure*}

Figure~\ref{level-cross-fig} shows the level-crossing diagrams corresponding 
to the $^2{\rm S}_{1/2}$ lower-level (top panels) and $^2{\rm P}_{3/2}$ 
upper-level (middle and bottom panels) of $^6$Li (panels (a), (b), (c)), 
$^7$Li (panels (d), (e), (f)), and Na (panels (g), (h), (i)). 
The incomplete PBE regime is characterized by crossing of magnetic substates 
belonging to different $F$ states, and also non-linear splitting of these 
magnetic components ($m$-states). In the case of the 
$^2{\rm S}_{1/2}$ lower-level, level-crossings are not found in the 
field strength range of 0--300\,G that we have considered. This validates 
our present approach, wherein we 
neglect level-crossing effects in the lower-levels, while the non-linear 
splitting is taken into account. In the case of $^2{\rm P}_{3/2}$ upper-level, 
the non-linear splitting and the crossing of magnetic substates of different 
$F$ states are clearly seen in the bottom panels of Fig.~\ref{level-cross-fig}. 
In particular, $^2$P$_{3/2}$ upper level of $^6$Li and $^7$Li exhibit 
respectively, 9 and 14 level-crossings for fields in the range 0--4\,G and 
0--10\,G, while that of Na\,{\sc i} exhibits 14 level-crossings for fields in 
the range 0--50\,G. For field strengths beyond the above-said range, the 
$m$-states start to bunch together indicating a gradual transition from 
incomplete PBE to the complete PBE regime. We recall that in the 
complete PBE regime the energy eigenvectors are of the form $|J\,I_s\,m_J\,
m_{I_s}\rangle$ and the HFS magnetic splitting varies linearly with the 
magnetic field strength. In the case of $^6$Li and $^7$Li 
the complete PBE regime is reached already for fields approximately equal to 
or larger than 50\,G (see panels (b) and (e)), while in the case of Na it 
is reached for $B\geqslant 200$\,G (see 
Fig.~\ref{level-cross-fig}(h)). 

We consider an isothermal planar model atmosphere, which is characterized by 
$(T,\Delta\lambda_{\rm D},\epsilon,r)$, where $T$ is total optical thickness 
along the atmospheric normal 
and $\Delta\lambda_{\rm D}$ is the Doppler width. In the present 
paper, $T$ is defined in terms of the frequency integrated line absorption 
coefficient. In a solar atmosphere, 
D$_2$ line of Li\,{\sc i} is optically thin, while that of Na\,{\sc i} is 
optically thick. To study the optically thin Li\,{\sc i} D$_2$ line, we
consider a self-emitting slab with parameters $T=10$,
$\Delta\lambda_{\rm D}=30$\,m\AA, $\epsilon=10^{-4}$, and $r=0$.
To study the optically thick Na\,{\sc i} D$_2$ line, we
consider a slab with parameters $T=10^7$, $\Delta\lambda_{\rm D}=25$\,m\AA,
$\epsilon=10^{-4}$, $r=10^{-7}$,
and a lower boundary condition of $I(\tau=T, x, {\bm n})=1$.
The magnetic field orientation is chosen as $\vartheta_B=90^\circ$ and
$\varphi_B=45^\circ$. We have considered field strengths between 
0 and 300\,G. To reduce the computational costs, we have 
considered angle-averaged magnetic PFR functions for all the results 
we have presented (although our code 
can also handle angle-dependent magnetic PFR functions). 

It is important to note that, although we have chosen the atomic 
parameters corresponding to the D$_2$ lines of Li\,{\sc i} and Na\,{\sc i}, 
the isothermal atmospheric models that we have considered for these two lines 
do not represent the corresponding solar conditions. This is particularly true 
in the case of Li\,{\sc i} D$_2$ line. In the solar case, the line-center 
opacity in this line is only slightly larger than that of the continuum 
due to the relatively low abundance of Li\,{\sc i} in the solar
atmosphere. Thus Li\,{\sc i} D$_2$ 
line forms at atmospheric heights corresponding to the photosphere, namely 
slightly above the $\tau_c=1$ layer for the continuum at 5000\,\AA. 
Furthermore, the line opacity is expected to be significantly smaller than 
that of the continuum outside the Doppler core. As a result the scattering 
polarization signal in this line is confined to the Doppler core, which 
quickly drops to the continuum polarization level when moving outside the 
core region. This can be clearly seen in the observations presented in 
\citet[][see also \citealt{jos11}]{skg00} as well as in the previous 
theoretical investigations based on single scattering 
\citep[see][]{blt09,snss15}. In contrast the results 
presented in Figs.~\ref{stokes-li6d2}--\ref{stokes-li76d2} show significant 
wing polarization signal even for ${\bm B}=0$. This is because our model 
atmosphere for Li\,{\sc i} D$_2$ line is a self-emitting slab of total optical 
thickness $T=10$. Also the continuum absorption coefficient is set to zero 
(i.e., the parameter $r=0$). {\textbf{Furthermore, we neglect the effects of elastic 
collisions, despite the fact that 
this line forms deep in the solar atmosphere, where 
elastic collision rates are high.}} Thus by neglecting the influence of 
continuum and elastic collisions, the amplitudes of the wing polarization of 
Li\,{\sc i} D$_2$ line presented in Figs.~\ref{stokes-li6d2}--\ref{stokes-li76d2} 
are overestimated. However, choice of such a model atmosphere, helps in 
clearly bringing out the signatures of different physical mechanisms discussed  
in this paper, that become much more apparent when PFR and transfer effects 
are included. Indeed the results presented in the present paper do not aim 
at an accurate modeling of the scattering polarization in the solar 
Li\,{\sc i} and Na\,{\sc i} D$_2$ lines. {\textbf{Instead, the goal of the 
theoretical calculations performed with isothermal slab atmospheres presented 
below is simply to highlight the signatures of incomplete PBE, PFR, and 
radiative transfer, among other effects.}} Hence the Li\,{\sc i} and 
Na\,{\sc i} D$_2$ lines considered 
in this paper are only ``theoretical model lines'' and do not represent 
actual solar lines, although we continue to refer to them as Li\,{\sc i} and 
Na\,{\sc i} D$_2$ lines for brevity.

\subsection{Theoretical Stokes Profiles of Li\,{\sc i} D$_2$ Line}
\begin{figure*}[!ht]
\centering
\includegraphics[scale=0.355]{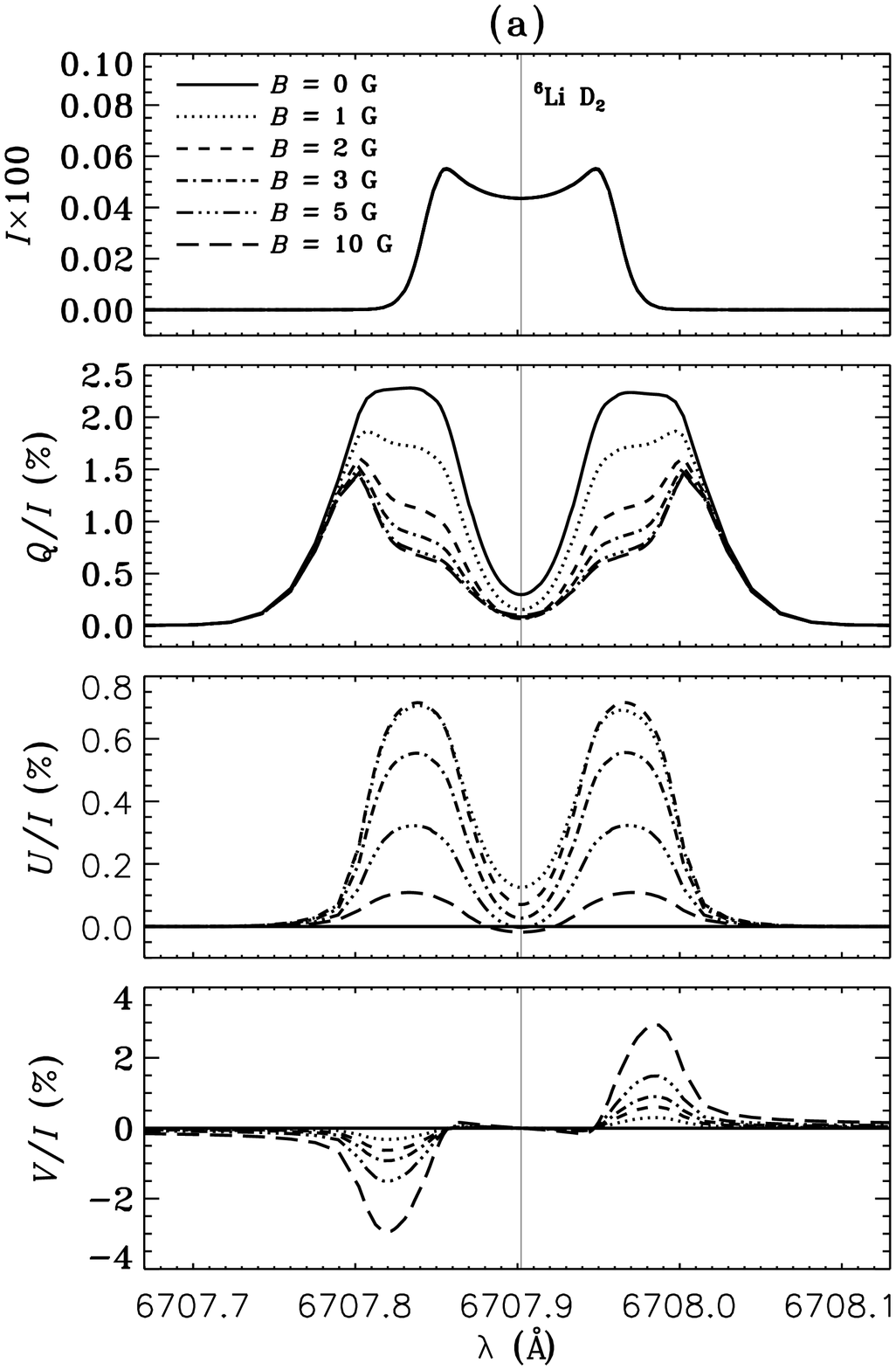} \ \ 
\includegraphics[scale=0.355]{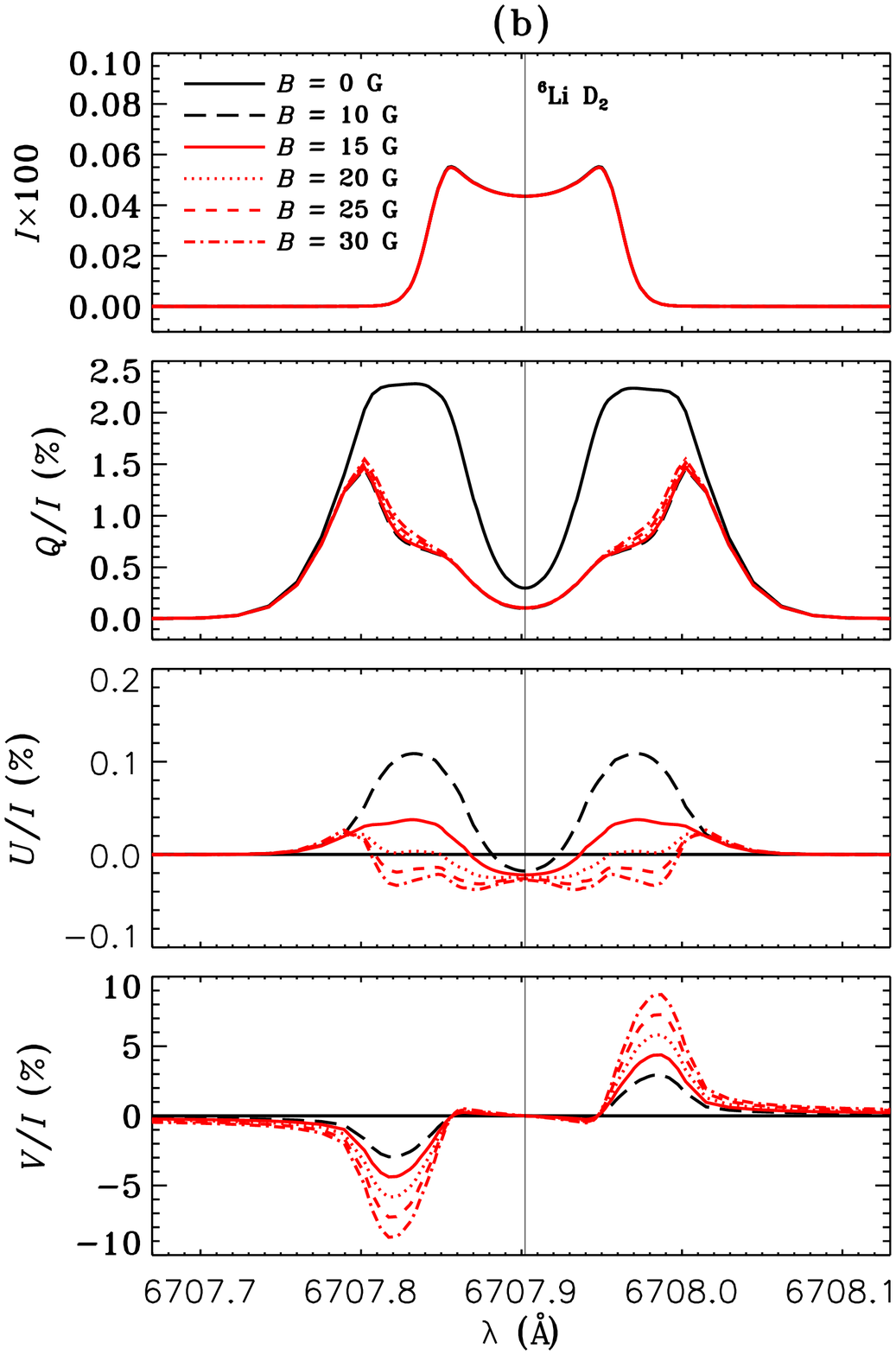} \ \ 
\includegraphics[scale=0.355]{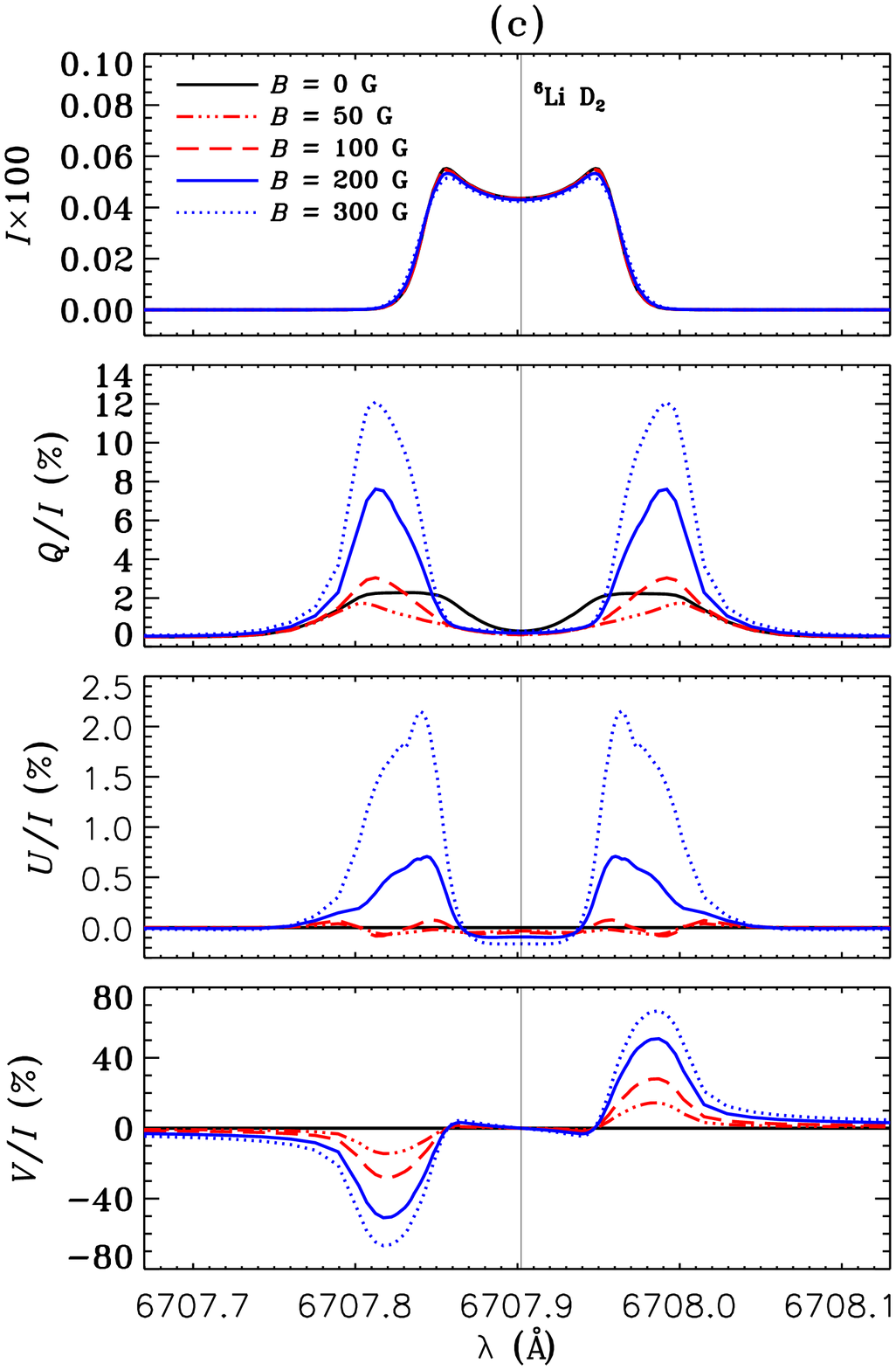}
\caption{The emergent $I$, $Q/I$, $U/I$, and $V/I$ profiles of a 
theoretical model line, whose atomic parameters correspond to those 
{\textbf{of the Li\,{\sc i} D$_2$ line, for which 100\%\ of the isotopes are 
$^6$Li. The line-of-sight is at $\mu=0.11$ and $\varphi=0^\circ$.}} 
Model parameters are $(T,\,\Delta\lambda_{\rm D},\,
\epsilon,\,r)=(10,\,30\,{\rm m}$\AA,$\,10^{-4},\,0)$. The magnetic field 
orientation $(\vartheta_B, \varphi_B) = (90^\circ,45^\circ)$. {\textbf{The 
field strength is varied between 0 and 300\,G.}}
}
\label{stokes-li6d2}
\end{figure*}

\begin{figure*}[!ht]
\centering
\includegraphics[scale=0.355]{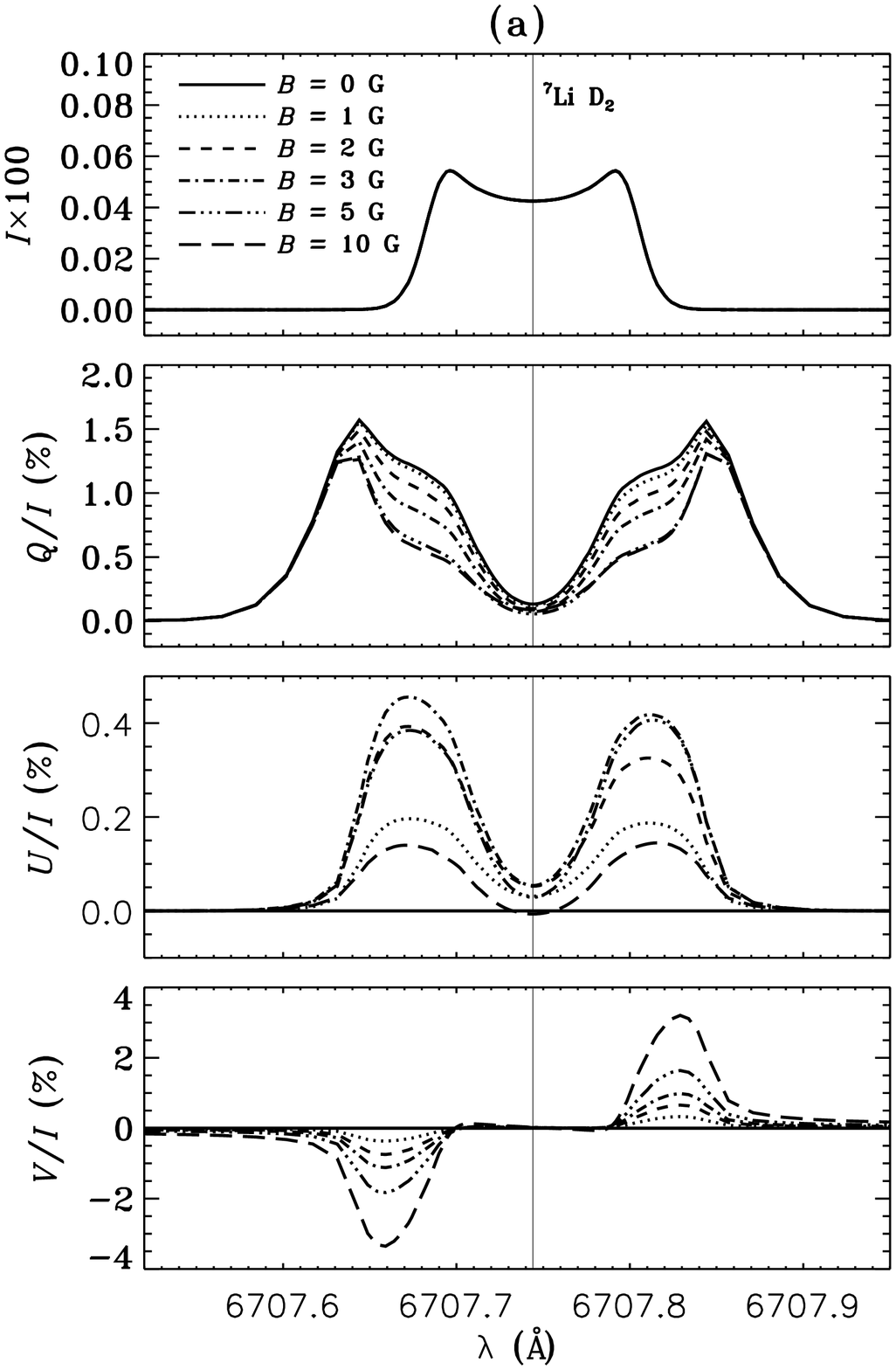} \ \ 
\includegraphics[scale=0.355]{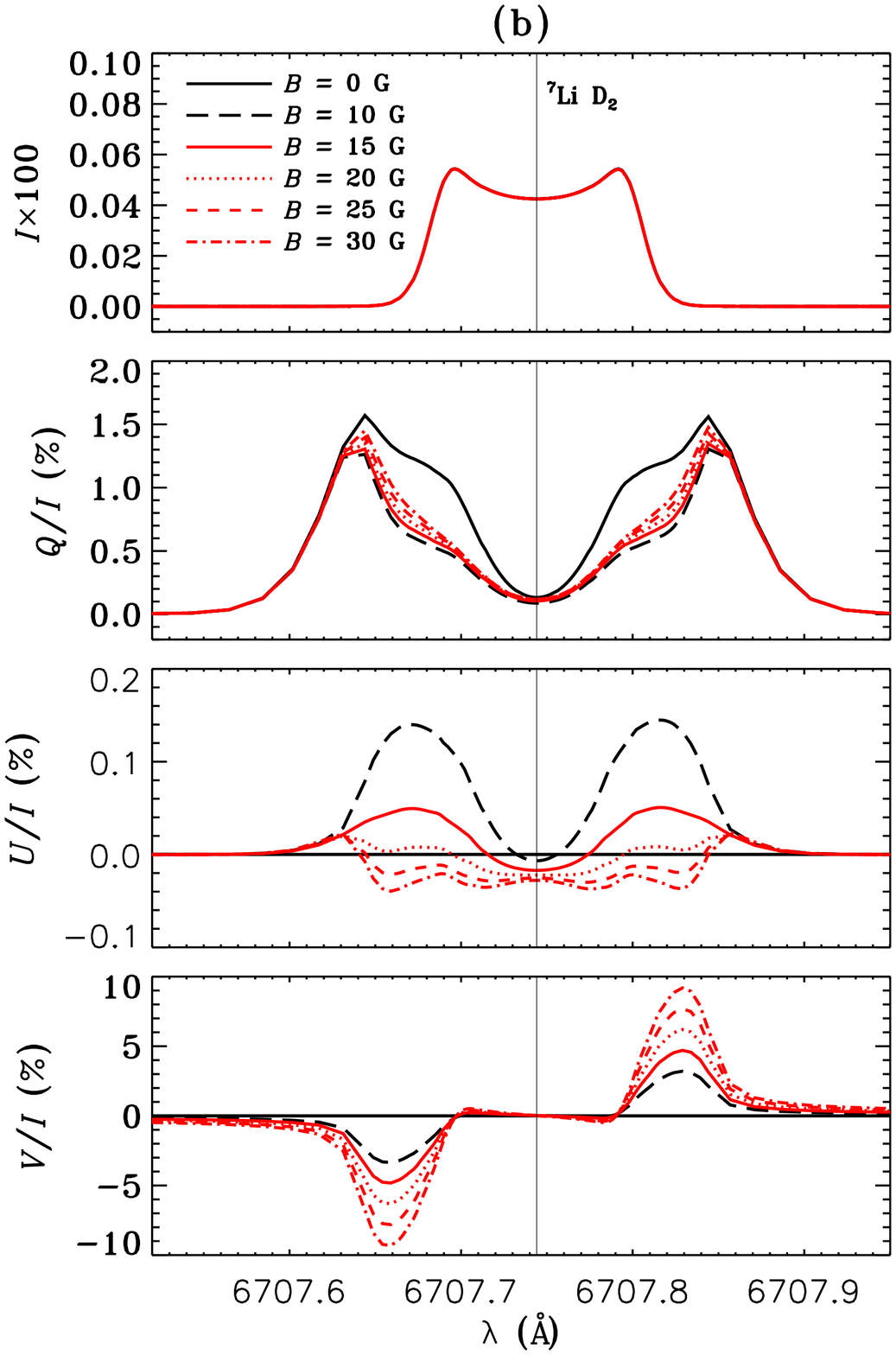} \ \ 
\includegraphics[scale=0.355]{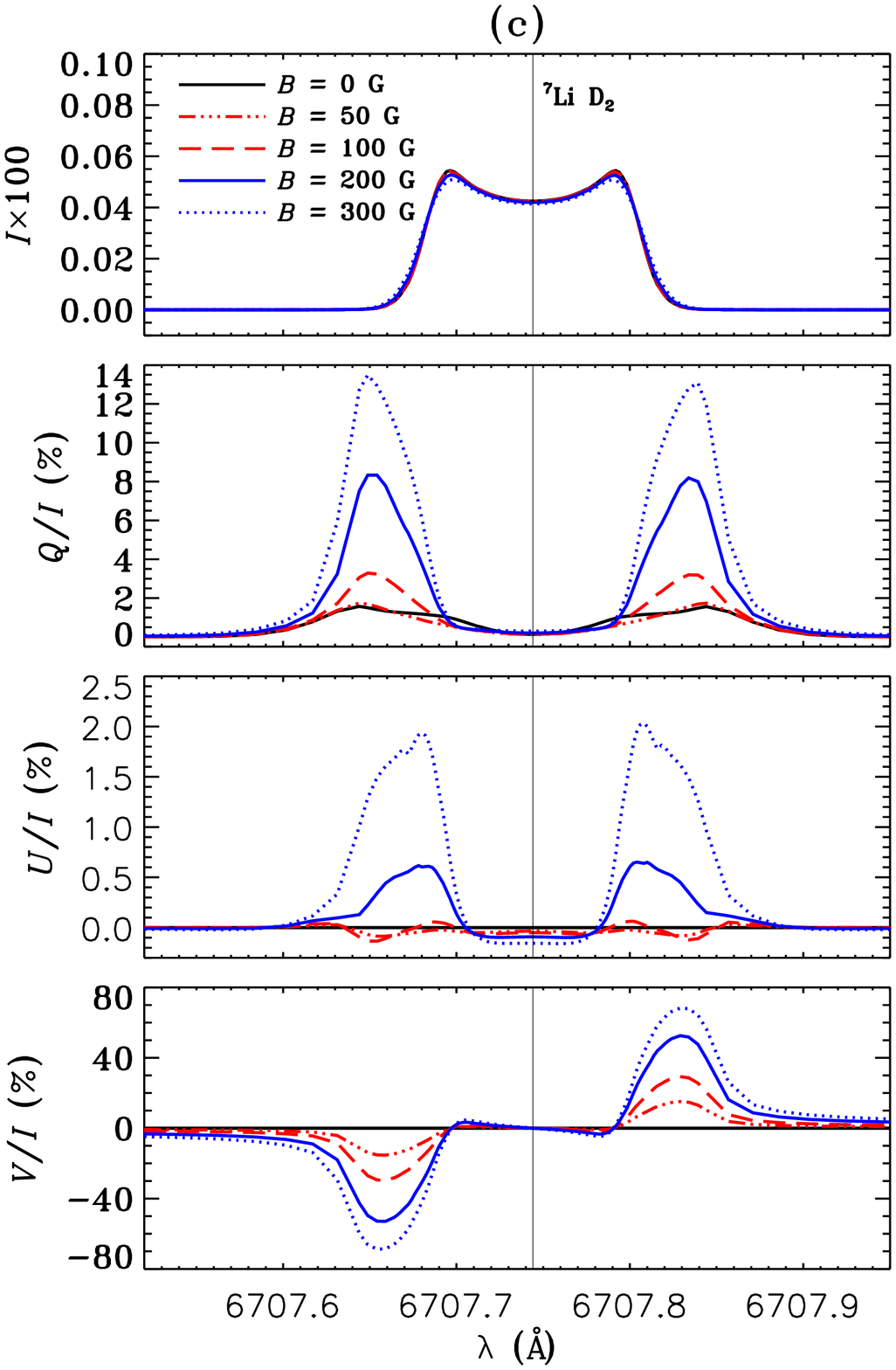}
\caption{The emergent $I$, $Q/I$, $U/I$, and $V/I$ profiles of a 
theoretical model line, whose atomic parameters correspond to those of 
{\textbf{the Li\,{\sc i} D$_2$ line, for which 100\%\ of the isotopes are 
$^7$Li. The line-of-sight is at $\mu=0.11$ and $\varphi=0^\circ$.}} 
Model parameters are $(T,\,\Delta\lambda_{\rm D},\,\epsilon,
\,r)=(10,\,30\,{\rm m}$\AA,$\,10^{-4},\,0)$. The magnetic field orientation
$(\vartheta_B, \varphi_B) = (90^\circ,45^\circ)$. {\textbf{The field 
strength is varied between 0 and 300\,G.}}
}
\label{stokes-li7d2}
\end{figure*}

\begin{figure*}[!ht]
\centering
\includegraphics[scale=0.355]{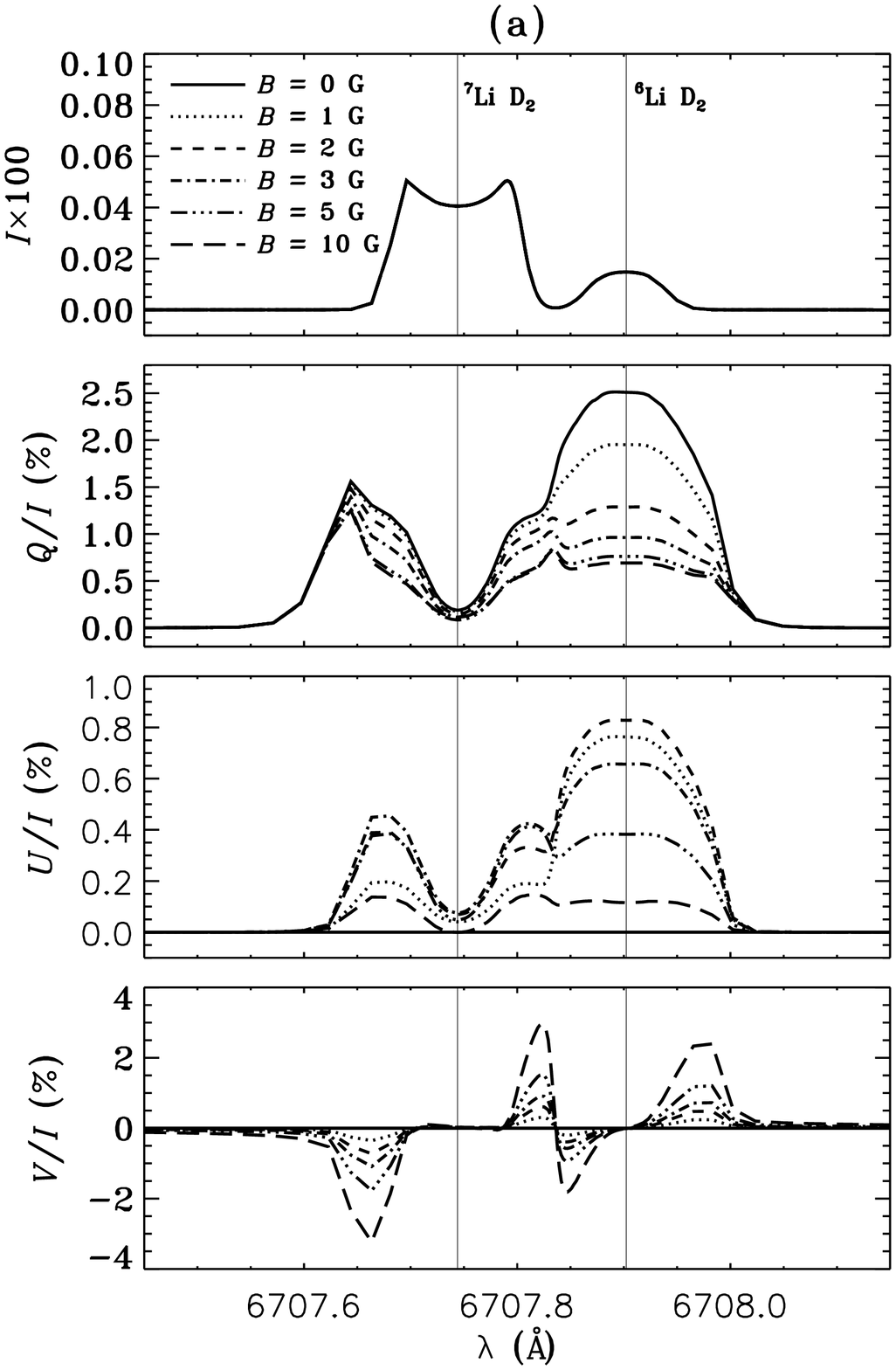} \ \ 
\includegraphics[scale=0.355]{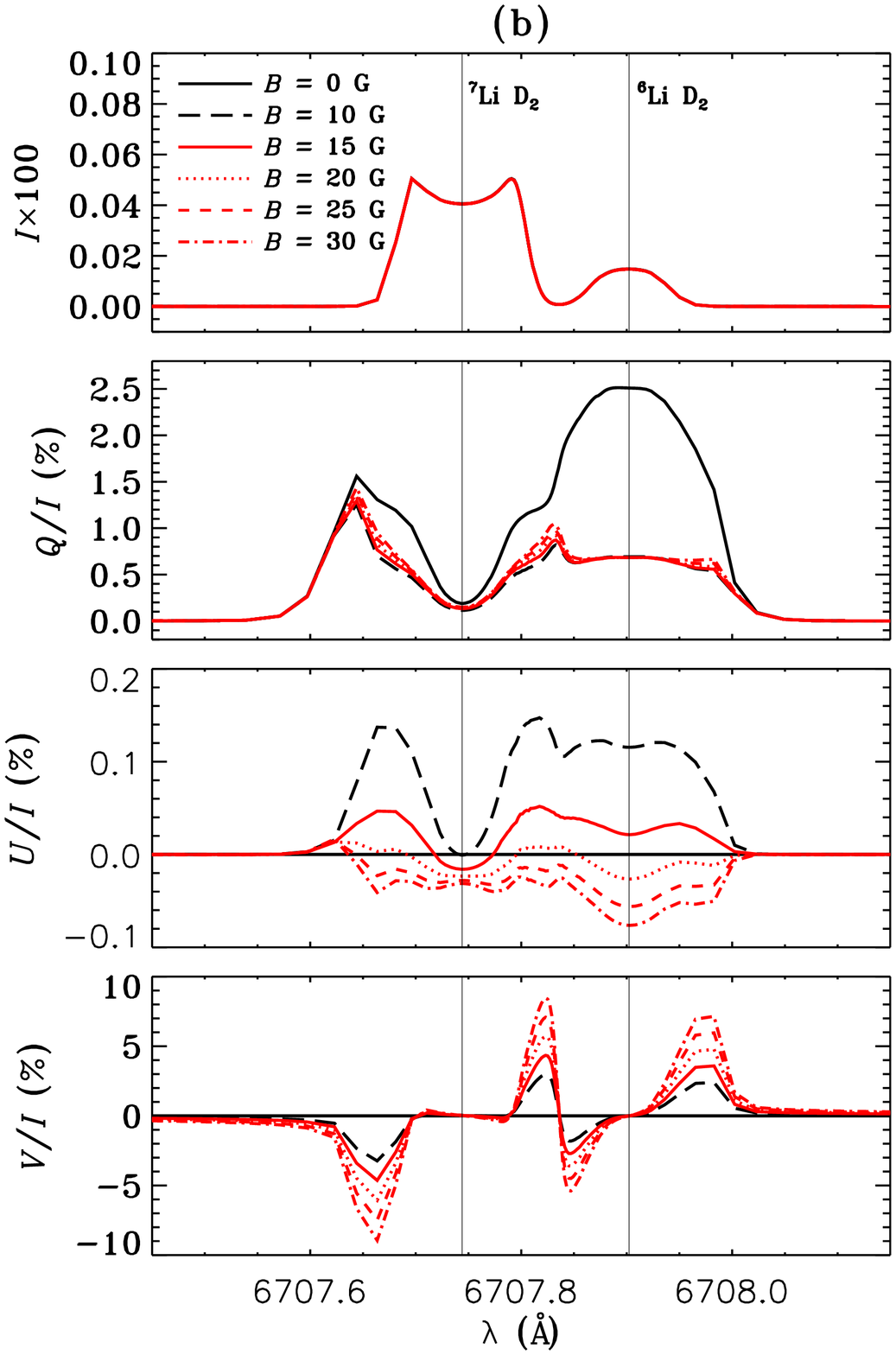} \ \ 
\includegraphics[scale=0.355]{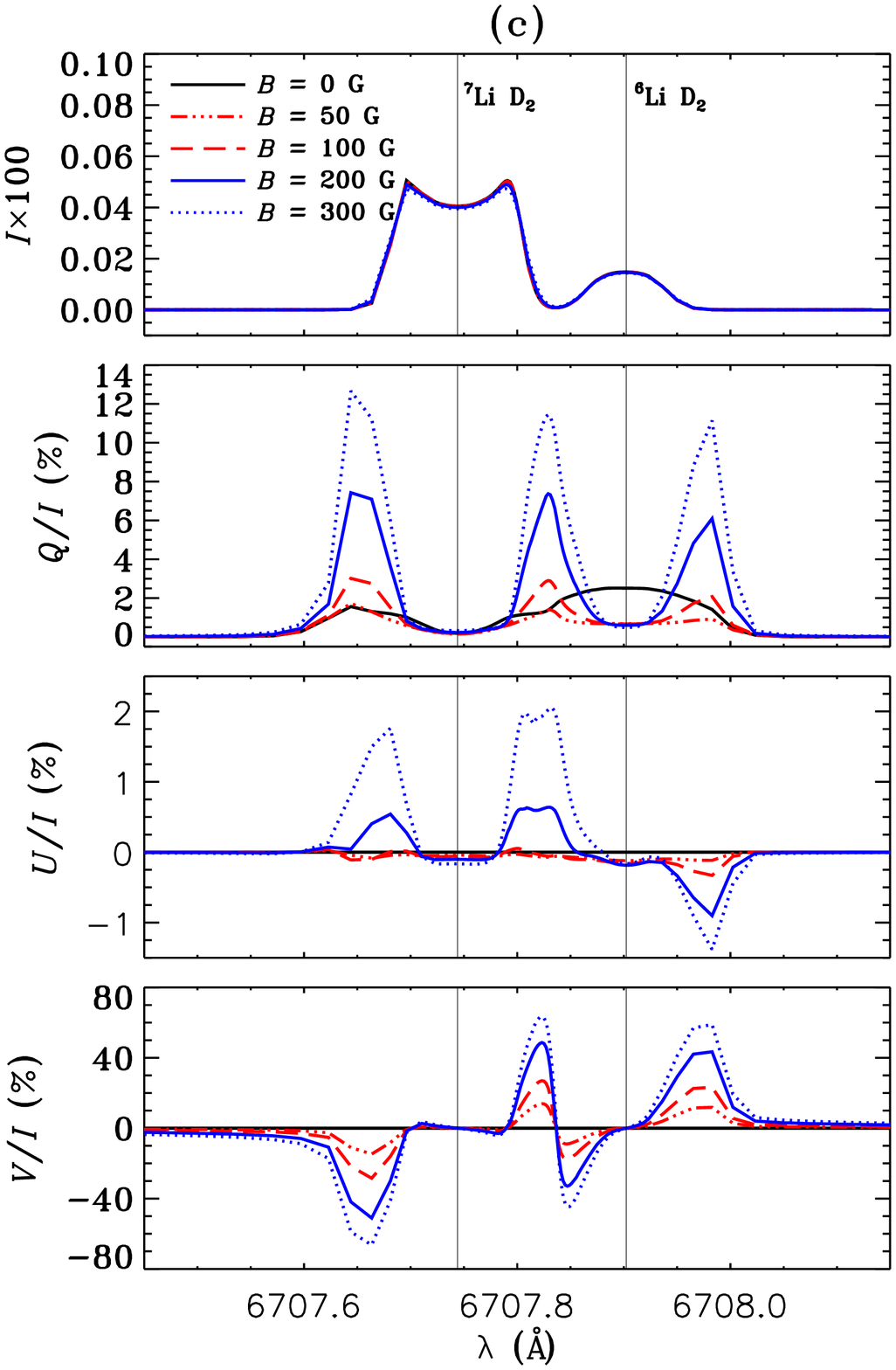}
\caption{The emergent $I$, $Q/I$, $U/I$, and $V/I$ profiles of $^7$Li and 
$^6$Li\,{\sc i} D$_2$ lines combined according to their isotopic abundance. 
Line-of-sight is at $\mu=0.11$ and $\varphi=0^\circ$.  
Model parameters are $(T,\,\Delta\lambda_{\rm D},\,\epsilon,
\,r)=(10,\,30\,{\rm m}$\AA,$\,10^{-4},\,0)$. The magnetic field orientation
$(\vartheta_B, \varphi_B) = (90^\circ,45^\circ)$. We recall that, 
these lines represent only theoretical model lines and not the 
actual solar lines. {\textbf{The field strength is varied between 0 and 
300\,G.}}
}
\label{stokes-li76d2}
\end{figure*}

Figures~\ref{stokes-li6d2}--\ref{stokes-li76d2} show the impact 
of magnetic fields of various strengths on the emergent Stokes profiles of 
a theoretical model line whose atomic parameters correspond to 
those of the Li\,{\sc i} D$_2$ 
line. Figure~\ref{stokes-li6d2} corresponds to the case of 100\%\ $^6$Li, 
Fig.~\ref{stokes-li7d2} to 100\%\ $^7$Li, and Fig.~\ref{stokes-li76d2} 
to the case of $^6$Li and $^7$Li weighted by their solar percentage 
abundance. Intensity shows a self-reversed emission profile, 
which is typical of an optically thin self-emitting slab. It remains 
insensitive to 
increase in $B$. This is because HFS magnetic splittings are too small 
compared to the Doppler width in the field strength range considered by us. 
Slight magnetic broadening can be seen in $I$ for 
$B=300$\,G (see e.g., blue dotted line in the $I$ panel of 
Fig.~\ref{stokes-li6d2}(c)). Hanle depolarization and rotation are seen in
$(Q/I, U/I)$ profiles for $B\leqslant 50$\,G (compare for example solid
black and red lines in Fig.~\ref{stokes-li6d2}(b)). For $B>50$\,G transverse 
Zeeman effect signatures are seen in $(Q/I, U/I)$ profiles. $V/I$ profiles 
show typical signatures of longitudinal Zeeman effect. 

In the case of $^7$Li, the non-linear splitting of HFS magnetic components 
produces a slight asymmetry about the line center in the $U/I$ profiles, 
for fields in the range $1\leqslant B < 10$\,G (see 
$U/I$ panel in Fig.~\ref{stokes-li7d2}(a)). For $1\leqslant B \leqslant 3$\,G, 
the blue side peak in $U/I$ is slightly higher than that on the red side. 
This behaviour is reversed for $B=5$\,G. For $B\geqslant 10$\,G, the two 
peaks are nearly symmetric. In the case of $^6$Li, this slight asymmetry is 
seen only for $B=1$\,G (see dotted line in $U/I$ panel of 
Fig.~\ref{stokes-li6d2}(a)). The $Q/I$ profiles of $^7$Li show Hanle 
depolarization until $B\leqslant 5$\,G (see $Q/I$ panel in 
Fig.~\ref{stokes-li7d2}(a)). For $5<B<50$\,G they tend to move towards the 
non-magnetic profile both in the line core as well as the wings (see $Q/I$ 
panel in Fig.~\ref{stokes-li7d2}(b)). This may be attributed to level-crossing 
effects, although they occur only in the range of 0 to 10\,G in 
the case of $^7$Li (see Fig.~\ref{level-cross-fig}(f)). In the case of $^6$Li 
this is seen clearly only in the wings of $Q/I$ profiles (see red lines in 
$Q/I$ panel of Fig.~\ref{stokes-li6d2}(b)). The level-crossing 
signatures are less pronounced in the case of 100\%\ $^6$Li than in 100\%\ 
$^7$Li line. This is because level-crossings for $^2$P$_{3/2}$ upper level 
of $^6$Li isotope occur at weaker fields than for the corresponding level 
{\textbf{of $^7$Li isotope, as the hyperfine structure splitting is smaller 
for $^6$Li isotope (see Table~\ref{atomic-parameters}).}} Level-crossing 
effects can also be seen in $Q/I$ profiles of both the isotopes of Li 
when combined according to their percentage abundance, particularly 
around the $^7$Li D$_2$ line position (see $Q/I$ panel in 
Fig.~\ref{stokes-li76d2}(b)). {\textbf{It is useful to note that, for 
$B>10$\,G the behavior of $Q/I$ profiles noted above could have a partial 
contribution from the transverse Zeeman effect, apart from level-crossings, 
deciphering the individual effects of these is indeed not trivial.}}

\begin{figure*}[!ht]
\centering
\includegraphics[scale=0.50]{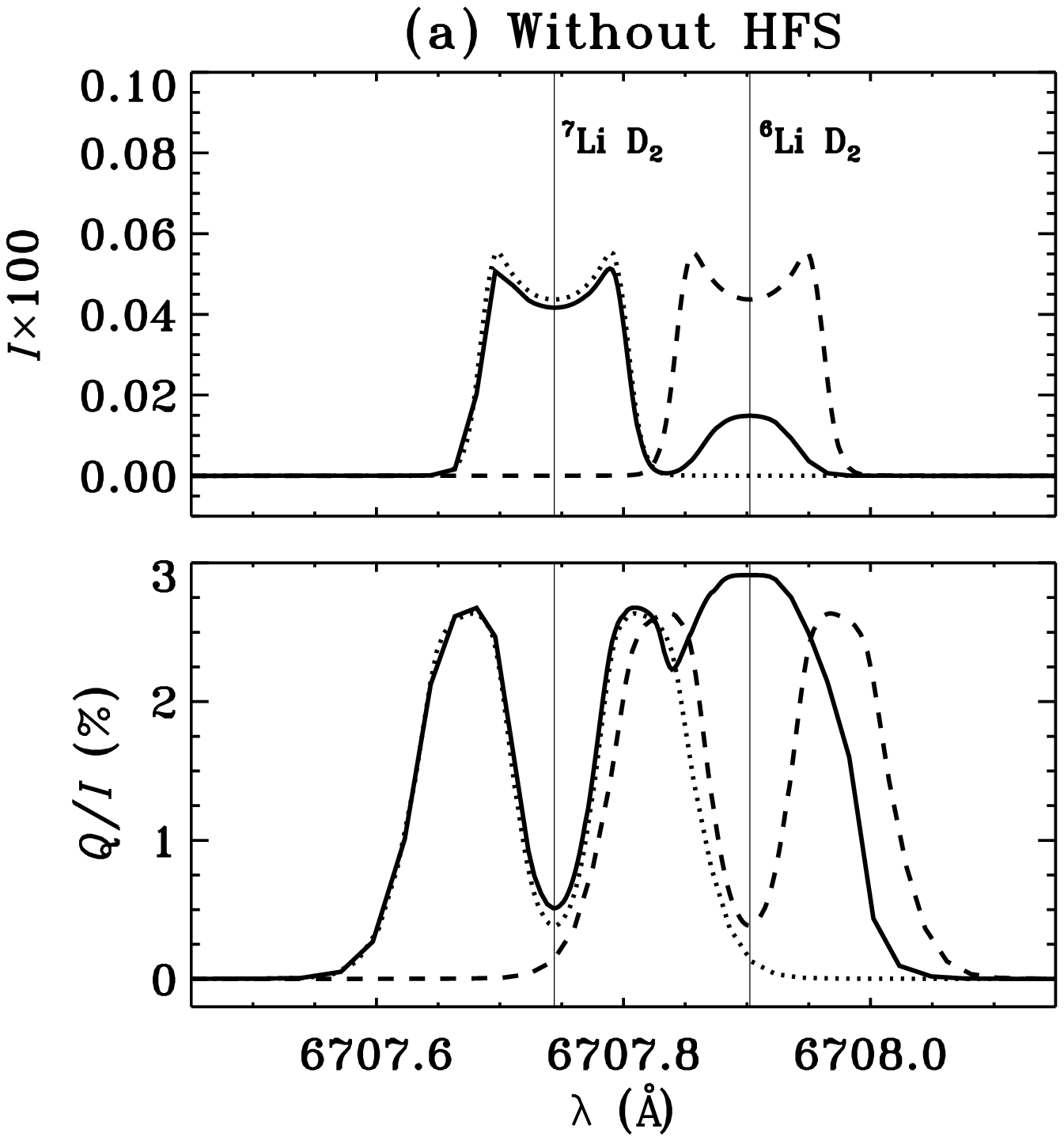} \ \ 
\includegraphics[scale=0.50]{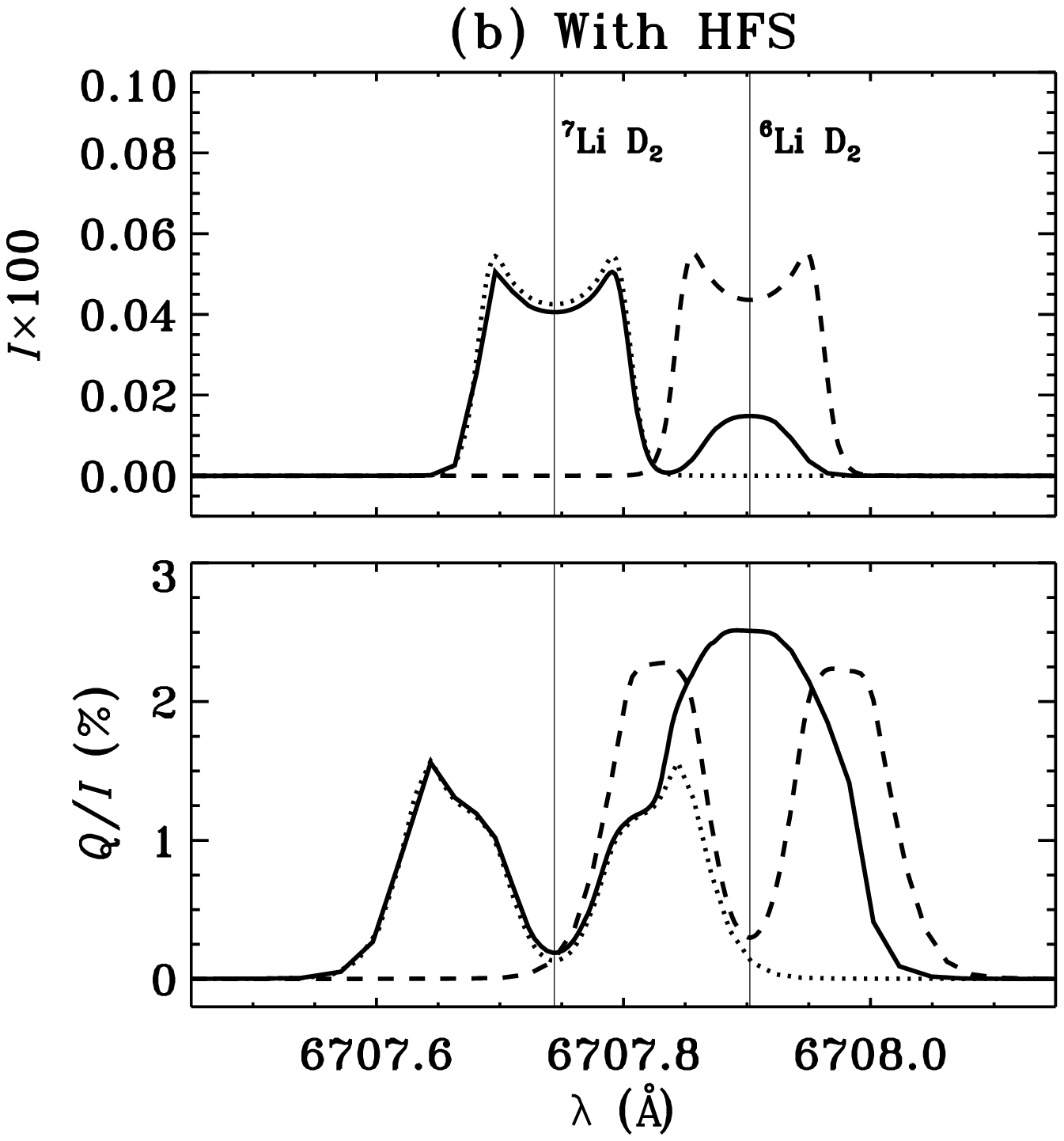}
\caption{A comparison of ($I$, $Q/I$) profiles computed with (panel (b)) and 
without (panel (a)) HFS in the non-magnetic case. Solid lines correspond 
to the combined case of $^6$Li and $^7$Li D$_2$ lines weighted by their 
percentage abundance, dotted lines to 100\,\% $^7$Li D$_2$ line, and dashed 
lines to 100\,\% $^6$Li D$_2$ line. Line-of-sight is 
at $\mu=0.11$ and $\varphi=0^\circ$. Model parameters are the 
same as in Figs.~\ref{stokes-li6d2}--\ref{stokes-li76d2}. 
}
\label{hfs-effect-fig}
\end{figure*}

Because of the isotope shift, the line center positions of the D$_2$ lines of 
$^6$Li and $^7$Li differ by about 160\,m\,\AA,\ which is 
considerably larger than the Doppler 
width of 30\,m\,\AA\ chosen by us. Thus in Fig.~\ref{stokes-li76d2} we see 
two distinct polarized line profiles in all the four Stokes parameters, 
apart from the blending in between the two lines modified by the radiative 
transfer effects. Because of the smaller isotopic abundance of $^6$Li, 
amplitude of $I$ at the position of $^6$Li D$_2$ line is smaller than that 
of $^7$Li. {\textbf{Furthermore, the self-reversal signature found in $I$ 
around $^6$Li 
D$_2$ line center disappears because of reduction in total optical thickness 
by a factor 0.0759\ in this line.}} However, in polarization, $^6$Li exhibits a 
signal which is comparable in magnitude to that of $^7$Li. In fact for 
$B=0$\,G, $Q/I$ at $^6$Li D$_2$ line center is larger than that at $^7$Li 
D$_2$ line center. This can also be seen clearly in the D$_2$ line profiles 
computed with 100\,\% $^7$Li and 100\,\% $^6$Li (compare black solid lines in 
Figs.~\ref{stokes-li6d2} and \ref{stokes-li7d2}). Indeed it is clearly visible 
in the single scattered profiles presented in 
\citet[][see the solid line in the right panels of their Fig.~3]{snss15}. 
This can be understood using the $[W_2(J_aI_sJ_b)]_{\rm hfs}$ factor for the 
case of two-level atom with HFS given in the un-numbered equation at the end of 
p.~617 of \citet{ll04}. Using that expression we find that 
$[W_2(J_aI_sJ_b)]_{\rm hfs}$ is $\approx 0.44$ for $^6$Li D$_2$ line and 
$\approx 0.25$ for $^7$Li D$_2$ line. Clearly $[W_2(^6{\rm Li})]_{\rm hfs} > 
[W_2(^7{\rm Li})]_{\rm hfs}$ causing larger $Q/I$ in $^6$Li D$_2$ line 
position than that at $^7$Li D$_2$. The transfer effects further modify 
this to produce the $Q/I$ line profile shown in Fig.~\ref{stokes-li76d2}. 
To further clarify the influence of HFS and the impact of the 
relative abundance of the two isotopes, we show in 
Fig.~\ref{hfs-effect-fig} a comparison of non-magnetic ($I$, $Q/I$) profiles 
computed with (panel (b)) and without (panel (a)) HFS. When HFS is neglected, 
the D$_2$ lines of both the isotopes have $W_2=0.5$. Thus $Q/I$ profiles of 
100\,\% $^7$Li and 100\,\% $^6$Li are identical except for an isotopic shift 
(compare dotted and dashed lines in $Q/I$ panel of 
Fig.~\ref{hfs-effect-fig}(a)). When the D$_2$ lines of these isotopes are 
combined according to their percentage abundance, the resulting $Q/I$ profile 
shows a peak at the $^6$Li D$_2$ line center. In the presence of HFS, the 
slightly larger $W_2$ for $^6$Li D$_2$ line results in a dominant peak as 
already discussed. 
\begin{figure*}[!ht]
\centering
\includegraphics[scale=0.40]{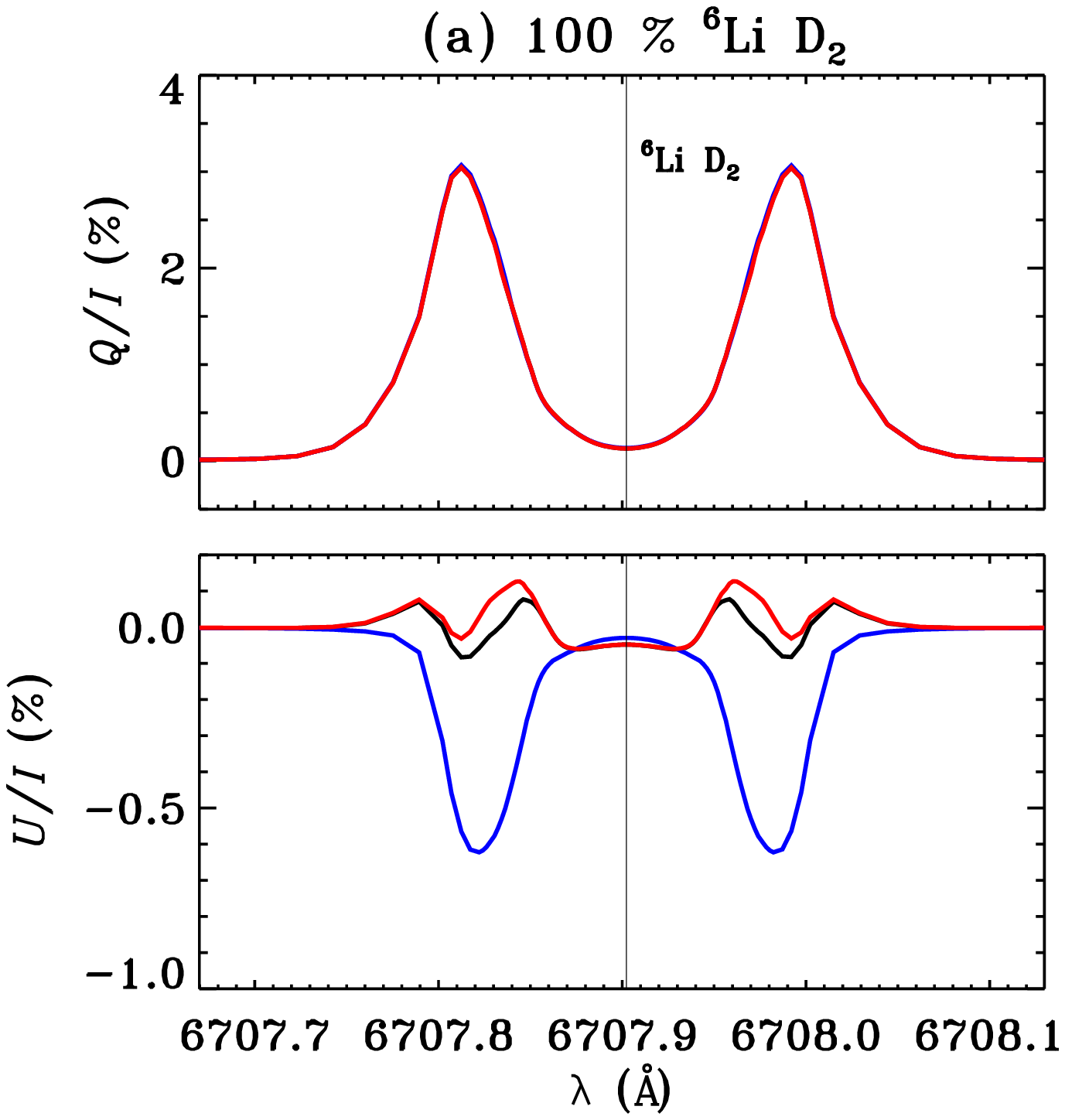} \ \ 
\includegraphics[scale=0.40]{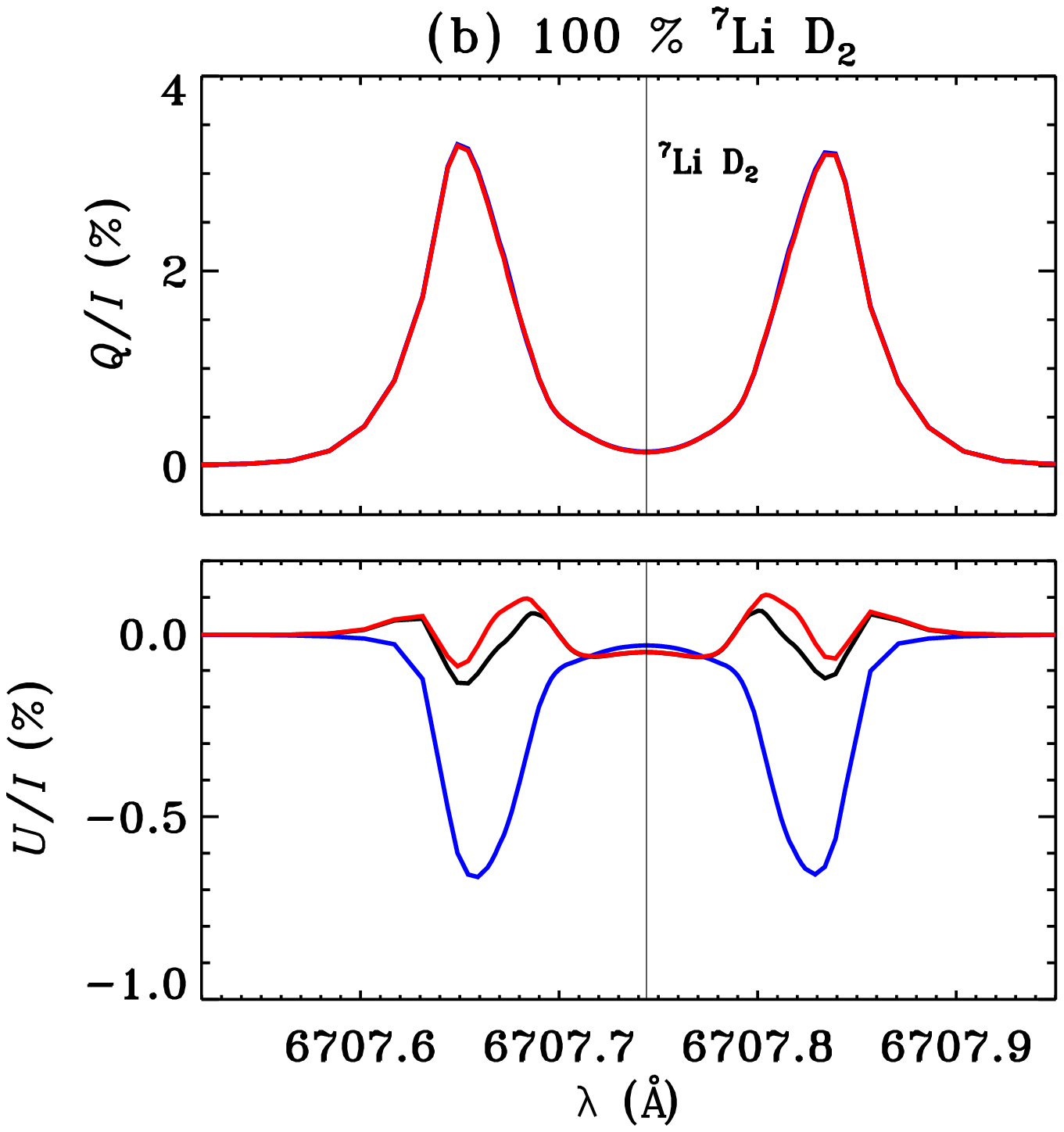}\ \ 
\includegraphics[scale=0.40]{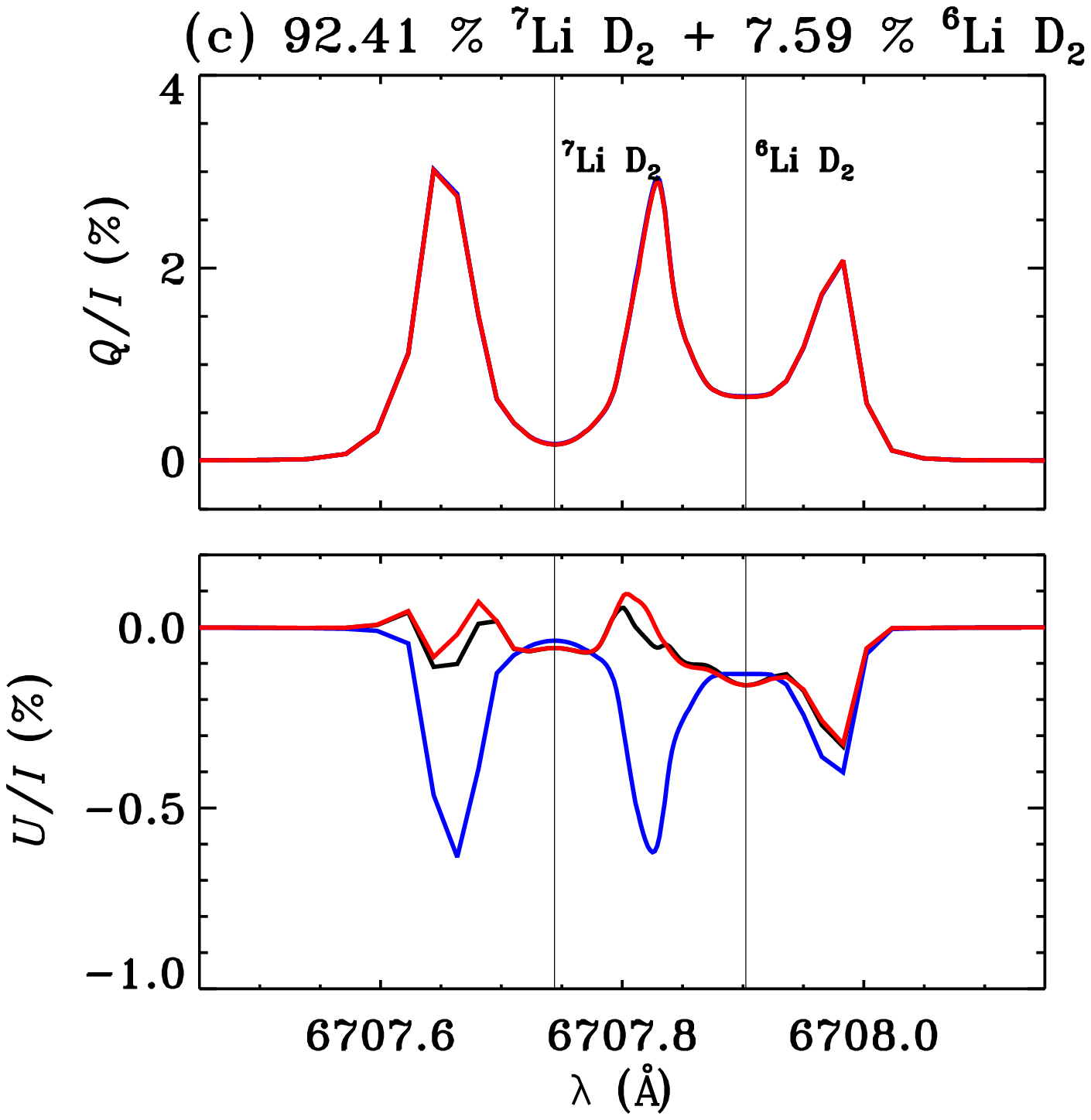}
\caption{A comparison of ($Q/I$, $U/I$) profiles computed including all the 
anomalous dispersion ($\chi_Q$, $\chi_U$, $\chi_V$) coefficients 
(black lines), computed with $\chi_V=0$ (blue lines), and with 
$\chi_Q=\chi_U=0$ (red lines) for $B=100$\,G.
Panel (a) corresponds to 100\,\% $^6$Li D$_2$ line, panel (b) to 100\,\% 
$^7$Li D$_2$ line, and panel (c) to the combined case of $^6$Li and $^7$Li 
D$_2$ lines weighted by their percentage abundance. Line-of-sight is 
at $\mu=0.11$ and $\varphi=0^\circ$. Model parameters are the 
same as in Figs.~\ref{stokes-li6d2}--\ref{stokes-li76d2}. All the three 
lines coincide in the $Q/I$ panels. 
}
\label{li76d2-qubyi-moeffect-b100}
\end{figure*}
\begin{figure*}[!ht]
\centering
\includegraphics[scale=0.40]{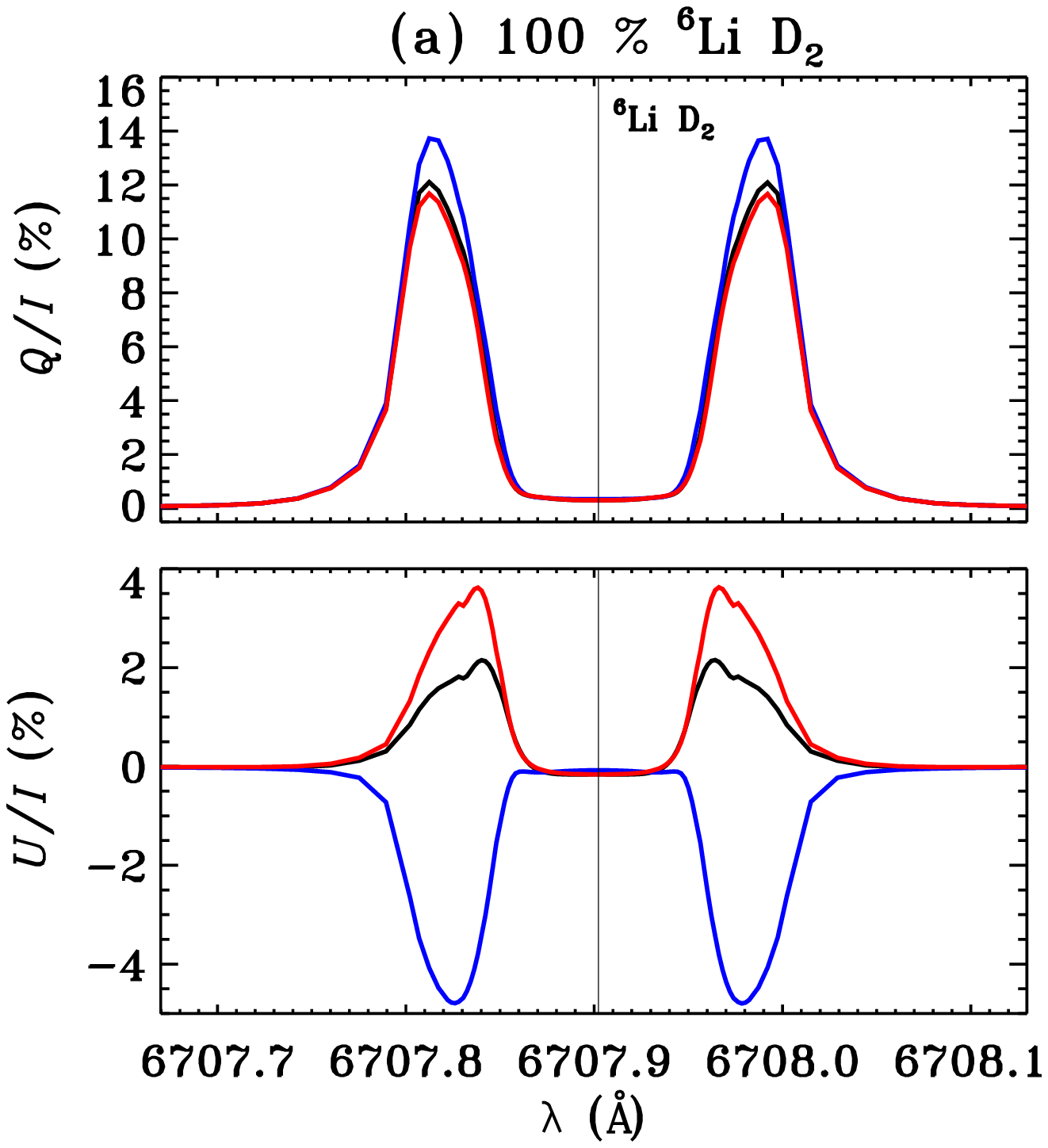} \ \ 
\includegraphics[scale=0.40]{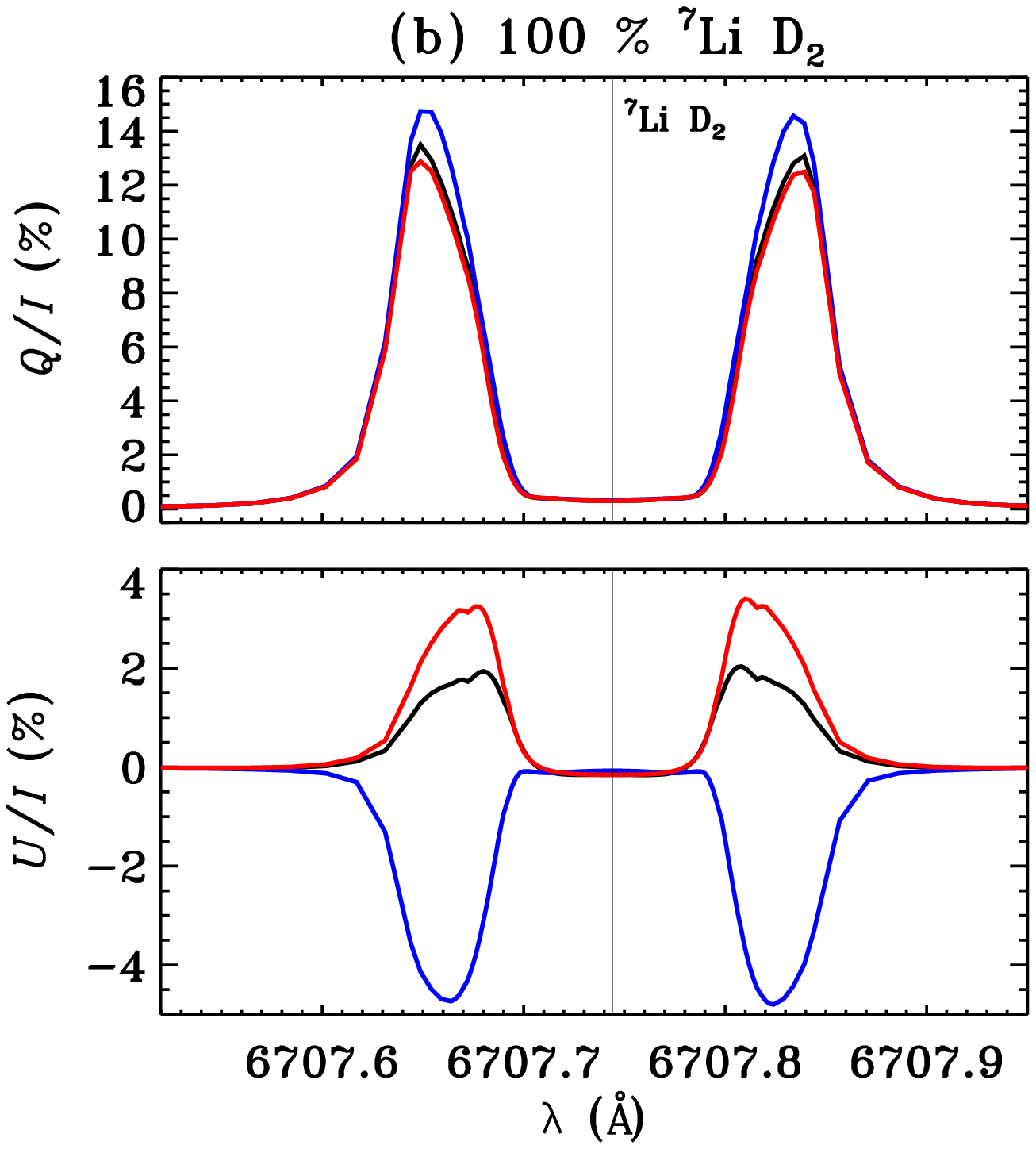}\ \ 
\includegraphics[scale=0.40]{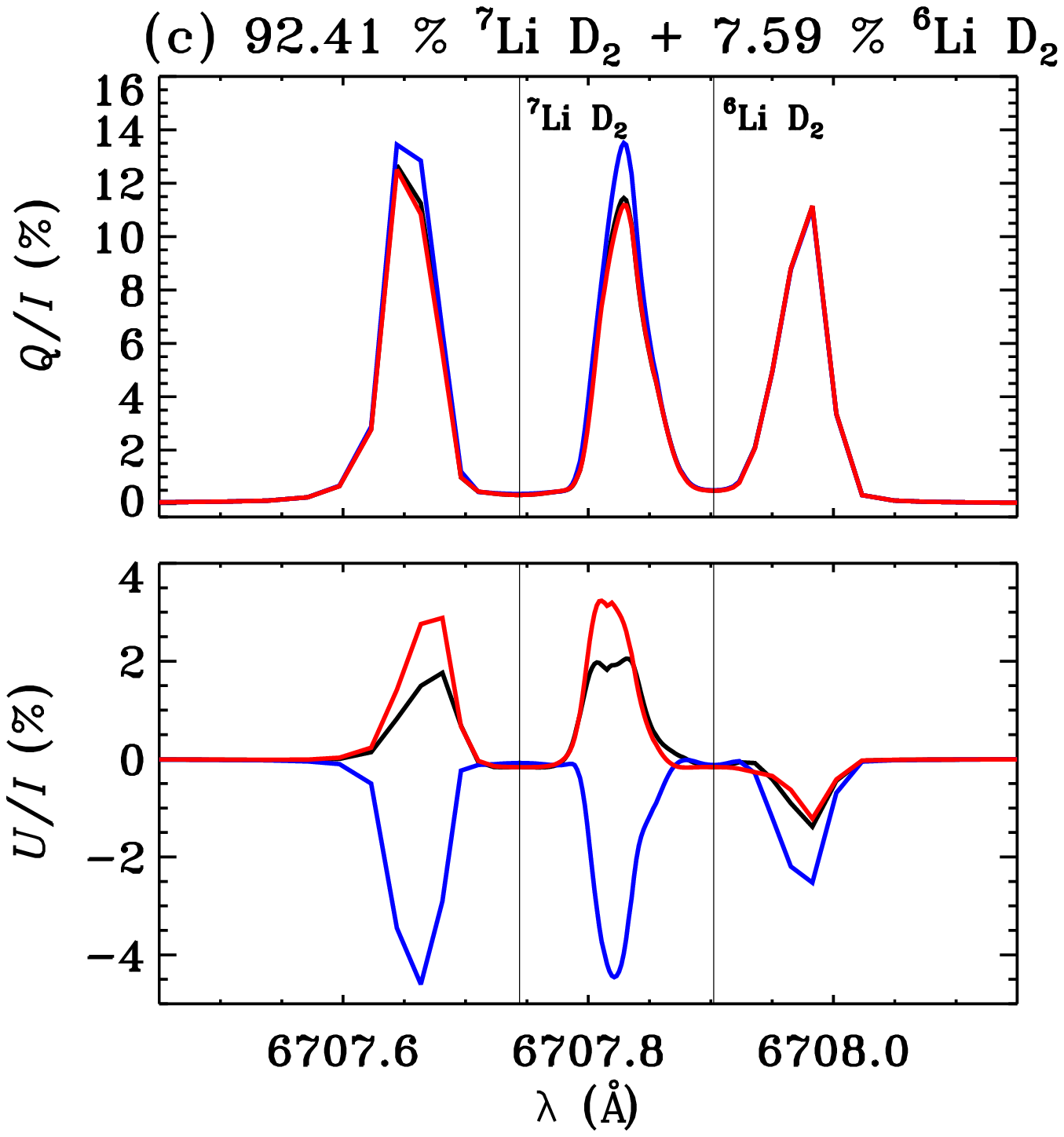}
\caption{Same as Fig.~\ref{li76d2-qubyi-moeffect-b100}, but for 
$B=300$\,G.
}
\label{li76d2-qubyi-moeffect-b300}
\end{figure*}

The $U/I$ profiles for $B\geqslant 50$\,G (see Fig.~\ref{stokes-li76d2}(c)) 
exhibit an interesting behaviour at the red wing 
position of $^6$Li D$_2$ line, namely that around that frequency 
$U/I$ changes sign, in 
contrast to the corresponding case of 100\,\% $^6$Li (compare for e.g., 
blue dotted lines in $U/I$ panels of Figs.~\ref{stokes-li6d2}(c) and 
\ref{stokes-li76d2}(c)). To understand this we compare in 
Figs.~\ref{li76d2-qubyi-moeffect-b100} and \ref{li76d2-qubyi-moeffect-b300} 
the ($Q/I$, $U/I$) profiles computed by including the full line absorption 
matrix ${\bf \Phi}$ (black lines), computed by setting only $\chi_V=0$ 
(blue lines), and only $\chi_Q=\chi_U=0$ (red lines) in ${\bf \Phi}$. From 
Eq.~(\ref{pbhfsrt-e2}) it is clear that anomalous dispersion coefficients 
$\chi_Q$, $\chi_U$, and $\chi_V$ couple the Stokes $Q$, $U$, and $V$. For the 
field strength regime considered in this paper only $\chi_V$ is significant, 
while $\chi_Q$ and $\chi_U$ are negligible for $B<50$\,G. It is well-known 
that the Zeeman effect gives rise to Faraday rotation (namely it 
rotates the plane of linear polarization and thereby converts $Q$ 
to $U$), which is described by $\chi_V$ \citep[see e.g.,][]{jos71,rmd89}. 
Zeeman effect also gives rise to Voigt effect (namely, it couples $V$ to $Q$ 
and $U$ and vice versa), which is described by 
$\chi_Q$ and $\chi_U$. The importance of Faraday 
rotation in the wings of ($Q/I$, $U/I$) profiles of optically thick lines has 
been demonstrated in \citet{abt16,abt17,abt18}, \citet{dcm16}, and 
\citet{sns17}. For the optically thin Li\,{\sc i} D$_2$ lines considered in 
this section, the Faraday rotation starts to become important in $U/I$ for 
$B\geqslant 30$\,G and in $Q/I$ for $B\geqslant 200$\,G, while the Voigt 
effect starts to show up in $U/I$ for $B\geqslant 50$\,G and in $Q/I$ for 
$B\geqslant 200$\,G. From Figs.~\ref{li76d2-qubyi-moeffect-b100} and 
\ref{li76d2-qubyi-moeffect-b300} we clearly see the rotation of plane of 
linear polarization due to the $\chi_V$ term (compare black and blue lines), 
which is larger for $B=300$\,G than for $B=100$\,G. The contributions of 
$\chi_U$ to $Q$ and $\chi_Q$ to $U$ are much smaller than that of $\chi_V$ 
(compare black and red lines). This is because $\chi_Q$ and $\chi_U$ are 
much smaller in magnitude than $\chi_V$ for the field strength regime 
considered here. Clearly, $\chi_U\,V$ tend to increase $Q/I$ slightly (for 
$B=300$\,G), while $\chi_Q\,V$ tend to reduce $U/I$ for the field geometry 
chosen here. For $B=300$\,G, $U/I$ profiles computed with  
100\,\% $^6$Li and $^7$Li are positive when the $\chi_V$ term is 
included and negative when it is neglected (compare black and 
blue lines in Figs.~\ref{li76d2-qubyi-moeffect-b300}(a) and 
\ref{li76d2-qubyi-moeffect-b300}(b), respectively). However when D$_2$ 
lines of $^6$Li and 
$^7$Li are combined according to their percentage abundance, the $\chi_V$ 
corresponding to $^6$Li is reduced by a factor $7.59$\,\% (which is 
the relative abundance of $^6$Li isotope), thereby reducing its impact 
on ($Q/I$, $U/I$) profiles at the red wing of $^6$Li D$_2$ line 
(see e.g., Fig.~\ref{li76d2-qubyi-moeffect-b300}(c)). Thus $U/I$ at the 
red wing of $^6$Li D$_2$ line continues to be negative in the 
combined case, in contrast to 100\,\% $^6$Li D$_2$ line case. 
{\textbf{Finally we note that, the impact of Faraday rotation and Voigt 
effect on the wings of $(Q/I, U/I)$ profiles of Li\,{\sc i} D$_2$ line are 
relatively smaller even for fields as large as $B=100$\,G (see 
Fig.~\ref{li76d2-qubyi-moeffect-b100}). This is because we have considered a 
self-emitting slab of $T=10$, wherein the transfer effects are highly reduced 
particularly in the wings. Faraday rotation and Voigt effect being 
propagation effects would exhibit larger influence on the wings of 
$(Q/I, U/I)$ profiles as the optical thickness of the medium increases. They 
modify the non-zero wing polarization in $Q/I$ produced by PFR. This is 
indeed the case for optically thick line considered in Section~\ref{nad2-sec}.}}

\subsection{Theoretical Stokes Profiles of Na\,{\sc i} D$_2$ Line}
\label{nad2-sec}
\begin{figure*}[!ht]
\centering
\includegraphics[scale=0.34]{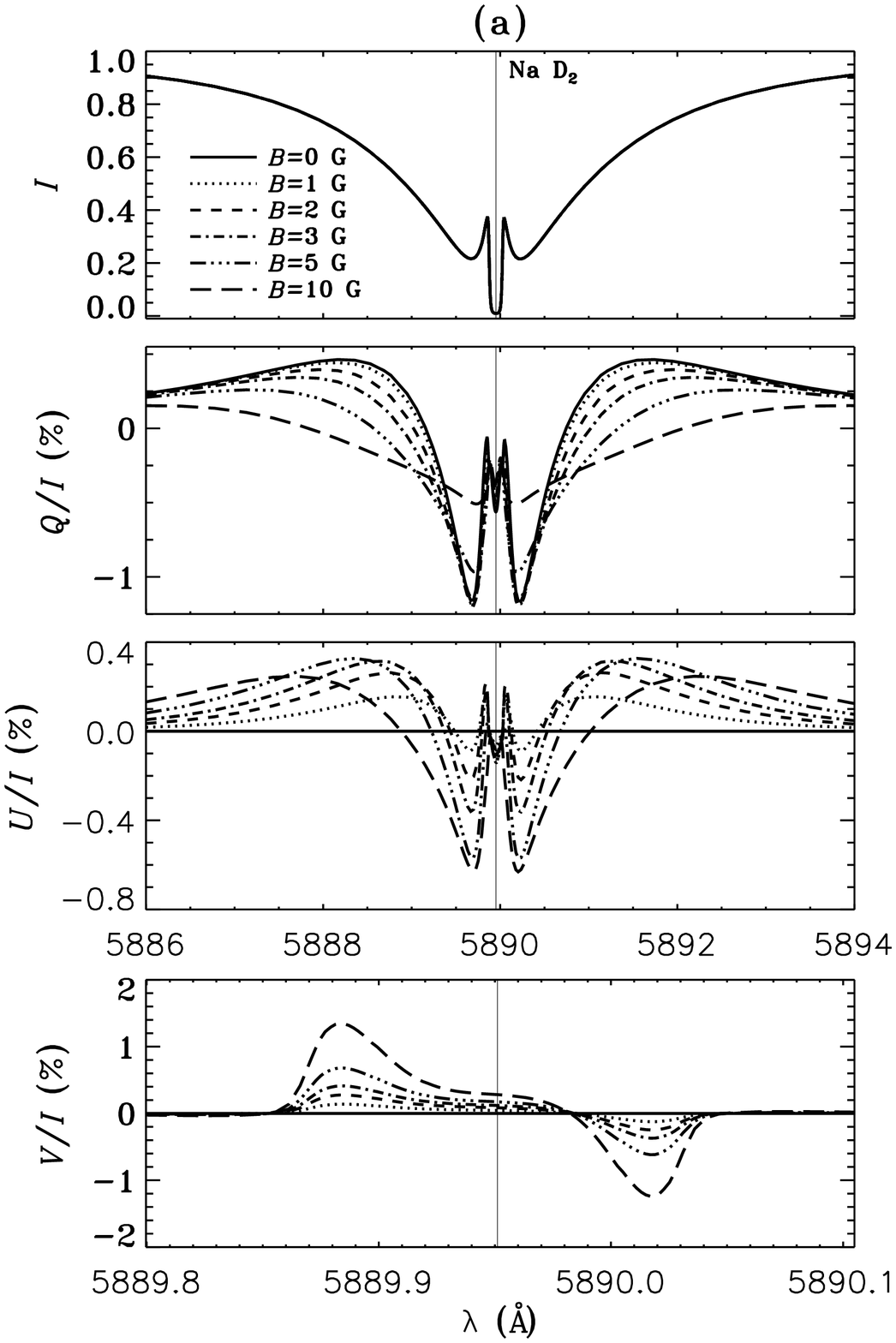}\ \ 
\includegraphics[scale=0.34]{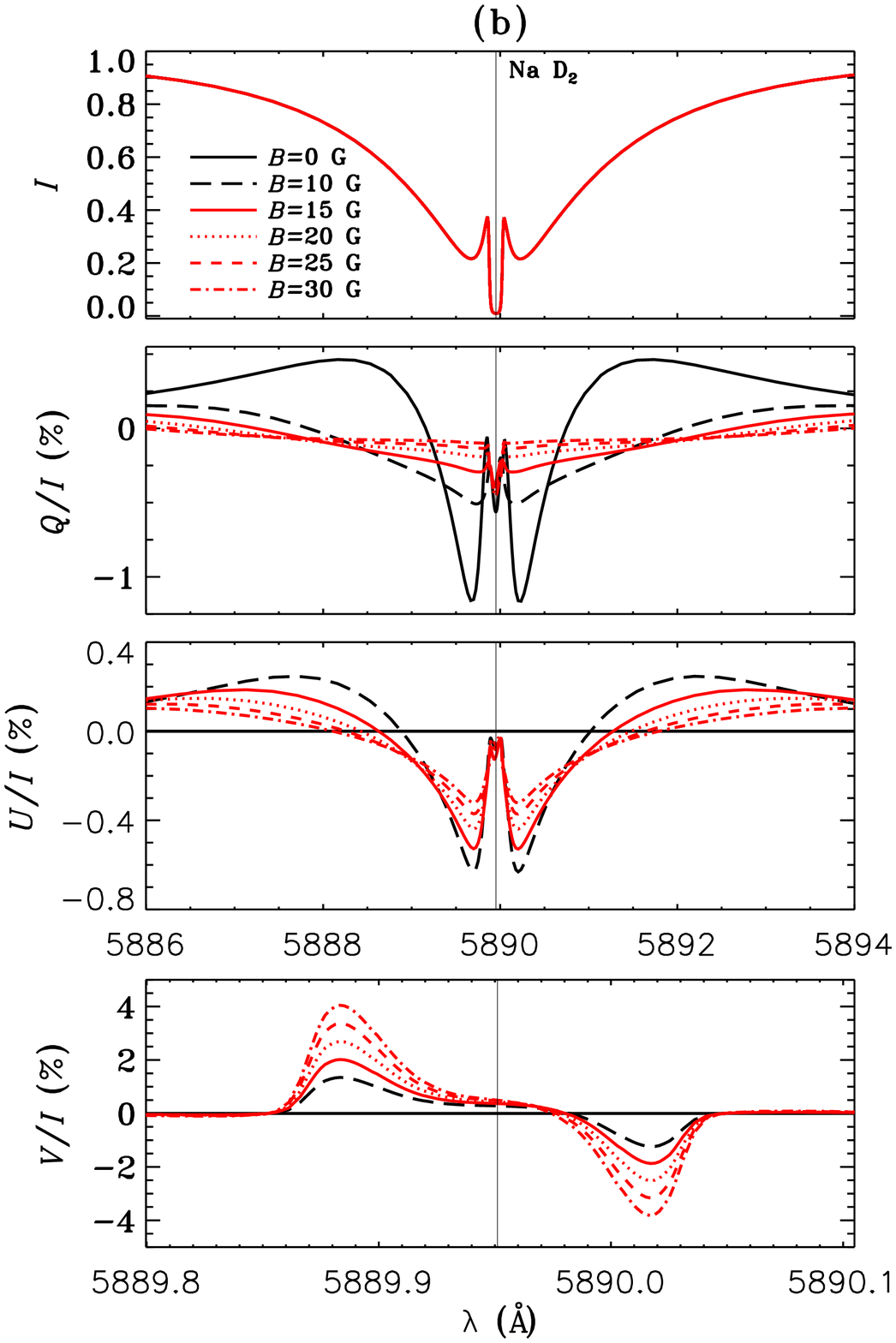}\ \ 
\includegraphics[scale=0.34]{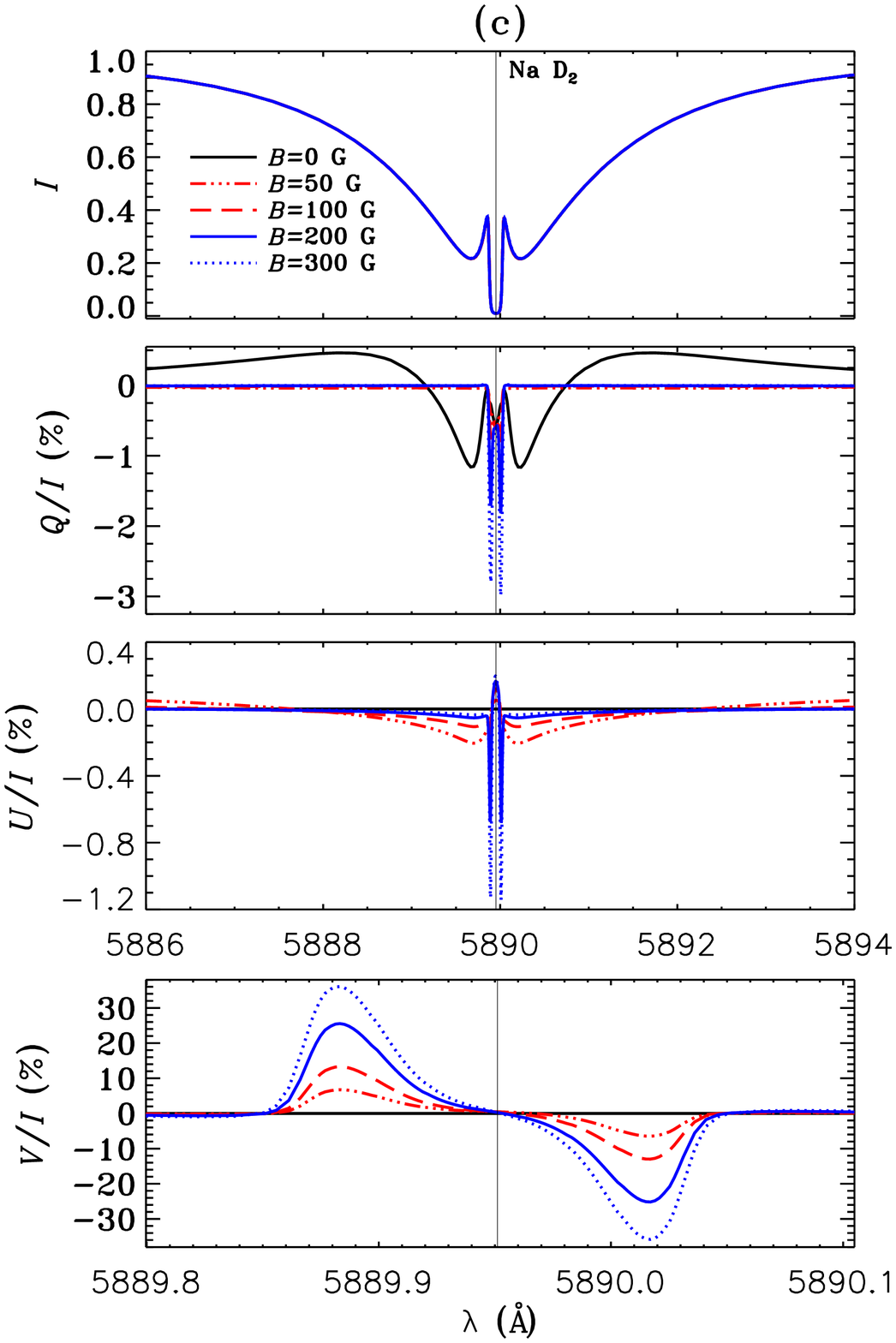}\\
\smallskip
\smallskip
\smallskip
\includegraphics[scale=0.38]{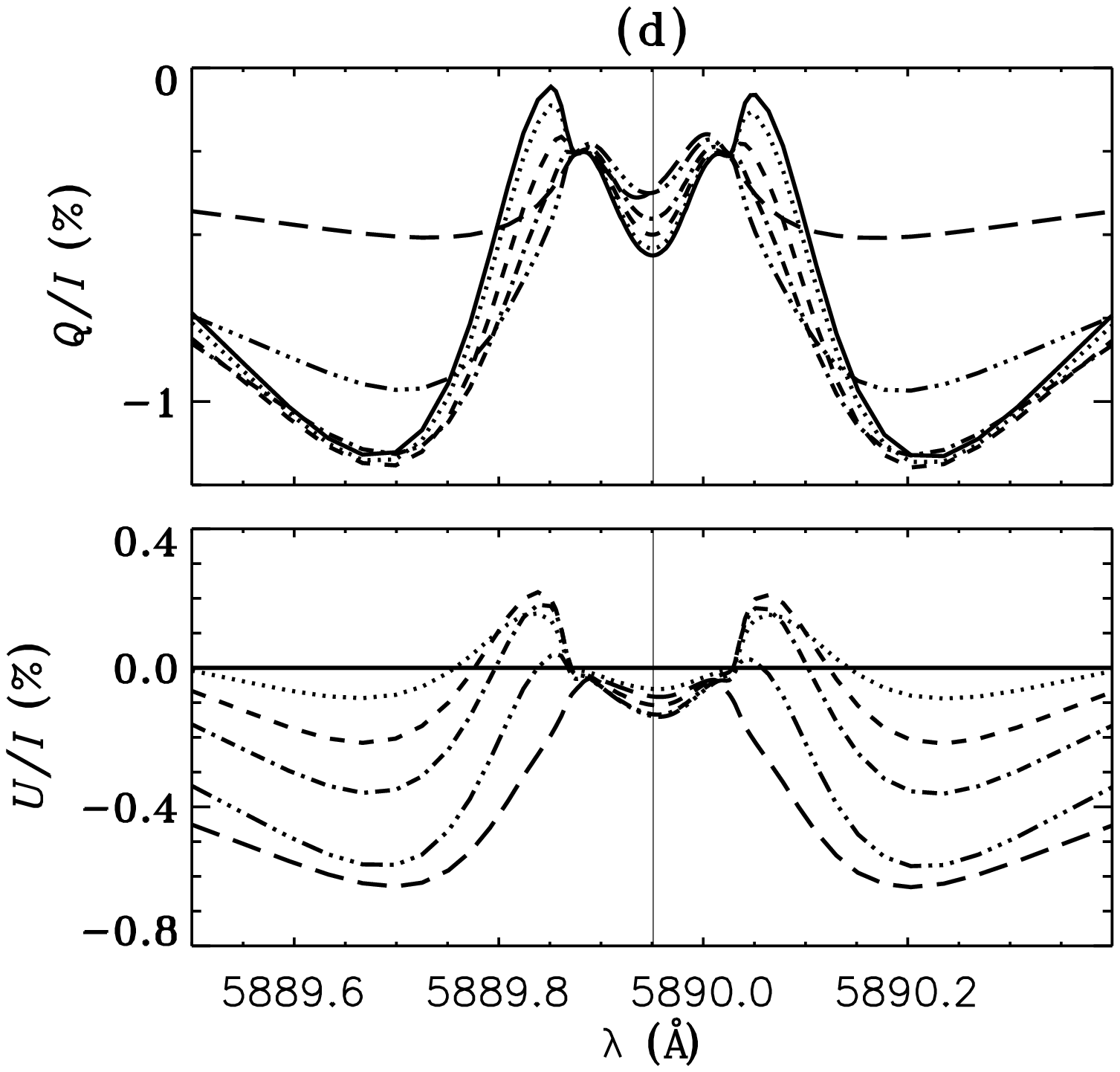}\ \ \ 
\includegraphics[scale=0.38]{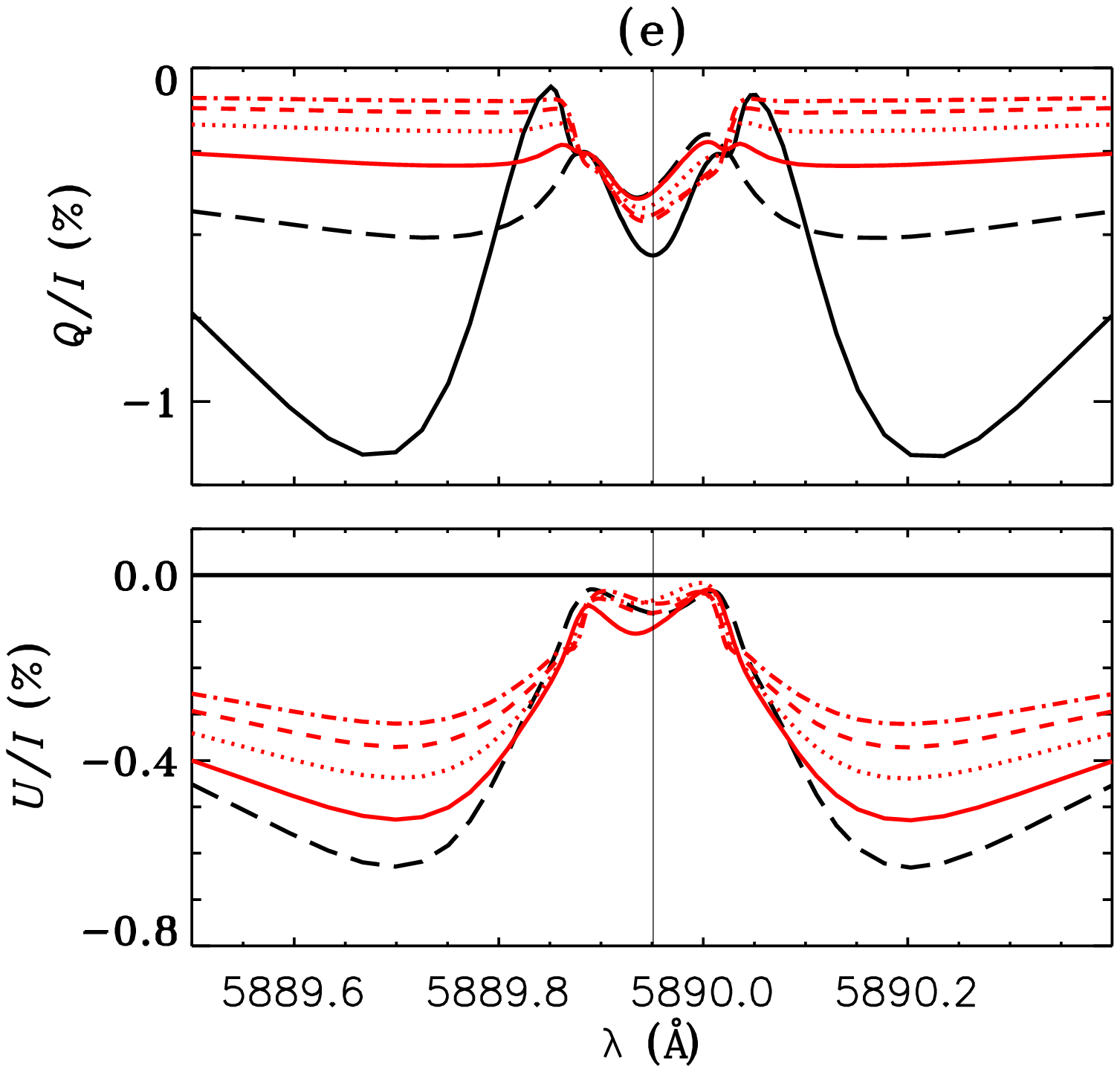}\ \ \ 
\includegraphics[scale=0.38]{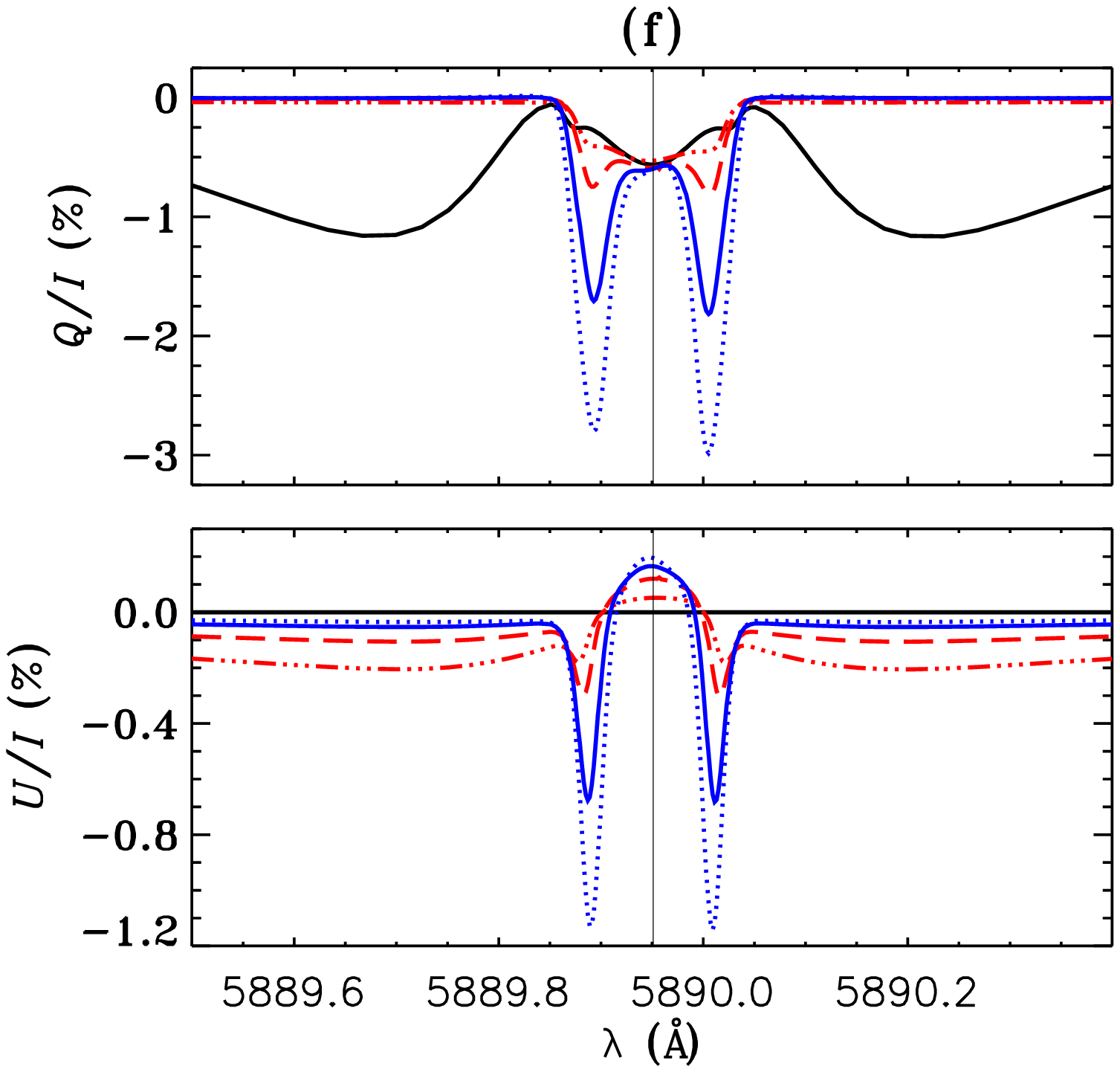}
\caption{The emergent $I$, $Q/I$, $U/I$, and $V/I$ profiles of 
a theoretical model line, whose atomic parameters correspond to those 
of the Na\,{\sc i} D$_2$ line, at $\mu=0.11$ and $\varphi=0^\circ$ for a 
range of field strengths. Model parameters are $(T,\,\Delta\lambda_{\rm D},\,
\epsilon,\,r)=(10^7,\,25\,{\rm m}$\AA,$\,10^{-4},\,10^{-7})$. The magnetic 
field orientation $(\vartheta_B, \varphi_B) = (90^\circ,45^\circ)$. Panels 
(d), (e), and (f) respectively show the magnified view in and around the line 
core region of $(Q/I, U/I)$ profiles shown in panels (a), (b), and 
(c). 
}
\label{stokes-nad2}
\end{figure*}

Figure~\ref{stokes-nad2} shows the impact of magnetic fields of 
various strengths on the emergent Stokes profiles of a 
theoretical model line whose atomic parameters correspond to those of 
the Na\,{\sc i} D$_2$ line. For the optically thick 
isothermal model that we have chosen for this line, we obtain an absorption 
line with minima in the near wings and also broad damping wings in 
$I$, which are due to PFR. In $Q/I$, PFR gives rise to a line core peak, 
followed by the core minima, near wing peaks, and far 
wing peaks for $B=0$ 
(see black solid line in $Q/I$ panels of Figs.~\ref{stokes-nad2}(a) and 
\ref{stokes-nad2}(d)). For $0<B<10$\,G, we see Hanle depolarization and 
rotation in the line core of ($Q/I$, $U/I$) profiles (see 
Fig.~\ref{stokes-nad2}(d)). For $10\leqslant B\leqslant 50$\,G, we see the 
signatures of level-crossing in the line core of ($Q/I$, $U/I$) profiles, 
namely they tend towards the non-magnetic value (see 
Fig.~\ref{stokes-nad2}(e)). For $B>50$\,G, transverse Zeeman effect signatures 
are seen in the line core of ($Q/I$, $U/I$) profiles (see 
Fig.~\ref{stokes-nad2}(f)). 
The Faraday rotation \citep{abt17,sns17}, which 
results in depolarization in the wings of $Q/I$ and generation of $U/I$ in 
the wings, strongly influence the wings of $(Q/I, U/I)$ profiles for the 
entire field strength regime considered here. In the case of 
Na\,{\sc i} D$_2$ line, the Voigt effect starts to show up in $U/I$ for 
$B\geqslant 100$\,G and in $Q/I$ for $B\geqslant 300$\,G, and its signatures 
are similar to those discussed in the case of Li\,{\sc i} D$_2$ lines 
(see Fig.~\ref{li76d2-qubyi-moeffect-b300}). The influence of 
Faraday rotation and Voigt effect on the Stokes profiles of Na\,{\sc i} D$_2$ 
line has been verified numerically by comparing the Stokes profiles computed 
with the full absorption matrix and those calculated by neglecting $\chi_V$, 
$\chi_Q$, and $\chi_U$ terms (like in Figs.~\ref{li76d2-qubyi-moeffect-b100} 
and \ref{li76d2-qubyi-moeffect-b300}). However, corresponding figures are not 
illustrated for brevity.

We note that, in the case of Na\,{\sc i} D$_2$ line, the upper 
$^2$P$_{3/2}$ level is in the incomplete PBE regime for $B<200$\,G, while 
the lower $^2$S$_{1/2}$ level continues to be in the incomplete PBE regime 
for $B> 200$\,G (see Figs.~\ref{level-cross-fig}(g), \ref{level-cross-fig}(h), 
and \ref{level-cross-fig}(f)). The signatures of incomplete PBE can be 
clearly seen in $V/I$ profiles, which are now asymmetric (see $V/I$ panel in 
Figs.~\ref{stokes-nad2}(a) and \ref{stokes-nad2}(b)). This is because of the 
asymmetric splitting of the magnetic components in the incomplete PBE regime. 
This results in a non-zero net circular polarization (namely $\int V 
d\lambda \ne 0$). The asymmetry in $V/I$ decreases as the field strength 
increases beyond 30\,G and the $V/I$ profiles become more and more 
anti-symmetric (see blue lines in the $V/I$ panel of 
Fig.~\ref{stokes-nad2}(c)). For more details on the relation between 
the incomplete PBE and the non-zero net circular polarization we refer the 
reader to \citet{ltc02}. It is interesting to note that the 
sigma components in the $Q/I$ profile for $B=200$ and $300$\,G (see blue 
lines in Fig.~\ref{stokes-nad2}(f)) are slightly asymmetric. This may be 
because the lower level of the D$_2$ line continues to be in the incomplete 
PBE regime.

\eject

\section{Conclusions}
\label{sec-conclu}
In the present paper we have considered the problem of polarized line 
formation in arbitrary magnetic fields taking into account PFR in scattering 
on a two-level atom with HFS. To numerically solve this problem, 
we have extended the scattering expansion method of \citet{fasn09}. 
Numerical solutions are presented by considering 
atomic systems representative of the D$_2$ lines of Li\,{\sc i} and 
Na\,{\sc i}. A range of field strengths from 0\,G to 300\,G are considered
for our studies. This field strength range covers both the incomplete 
and complete PBE regimes for both Li\,{\sc i} D$_2$ and Na\,{\sc i} D$_2$ 
lines (see Fig.~\ref{level-cross-fig}). Apart from the well-known signatures of 
Hanle and Zeeman effects, we have identified the signatures of incomplete 
PBE (namely, level-crossing, non-linear and asymmetric splitting), Faraday 
rotation and Voigt effects, and PFR in the Stokes profiles of the D$_2$ lines 
mentioned above, that are formed in an isothermal planar atmosphere. When 
compared to the single scattered Stokes profiles 
\citep[see for e.g.,][]{snss14,snss15}, the magnitude of the signatures of 
incomplete PBE regime is smaller, when radiative transfer effects are taken 
into account. However, radiative transfer effects need to be taken into 
account to accurately model the optically thick lines such as the Na\,{\sc i} 
D$_2$ line, wherein PFR manifests itself by producing wing PFR peaks in 
$Q/I$ profile, apart from the core peak. Although the signatures of 
incomplete PBE 
appear to be relatively less pronounced in the Stokes profiles formed in 
atmospheres where radiative transfer effects are important (in 
comparison with those formed in a single scattering event), it is essential 
to take them into account for an accurate determination of the magnetic fields 
using the D$_2$ lines of Li\,{\sc i} and Na\,{\sc i} considered in this 
paper and for other lines for which $I_s$ is non-zero. 
Thus, the incomplete PBE together with Hanle and Zeeman effects must 
be accounted for in order to reliably use spectropolarimetric measurements of 
these lines as a diagnostic tool to detect the solar/ stellar magnetic 
fields. 

We remark that in the present paper, for numerical simplicity, 
we have considered the academic case of isothermal planar atmospheres. 
To quantitatively evaluate the importance of incomplete PBE 
to determine solar/ stellar magnetic fields from the 
spectropolarimetric measurements of Li\,{\sc i} and Na\,{\sc i} D$_2$ lines, 
it is essential to 
generalize the present approach to handle realistic solar/ stellar atmospheres. 
However, at present this is computationally prohibitive. This is because, for 
an isothermal atmosphere itself, the computing time for a given magnetic field 
configuration (taking a single value of $B$, $\vartheta_B$, and $\varphi_B$) 
is on the order of few days to one or two weeks depending on the optical 
thickness of the atmosphere. The main memory required is also substantial, 
being on the order of 40 to 100\,GB. Although the requirement on the main 
memory can be somewhat reduced by avoiding storing the arrays in the main 
memory and using a secondary storage, reducing the CPU time requires 
parallelization of the code. Given this it is clear that there is still 
a long way to go before the present work can have practical applications 
as a good diagnostic tool. However the present paper certainly represents 
a first step in this direction. Finally, we note that, it may still be 
computationally feasible to apply the present work to realistic modeling. 
This may be achieved based on the last scattering approximation 
\citep{jos82}, a sophisticated version of which has been presented in 
\citet{lsaetal10} for the non-magnetic case, who show that it nearly 
reproduces the radiative transfer solution even for optically thick lines. 
An attempt to generalize this so called LSA-3 approach of \citet{lsaetal10} 
to include arbitrary fields has been presented in \citet{ks16}.

\acknowledgments
We acknowledge the use of the high-performance computing facility at 
Indian Institute of Astrophysics. We thank the referee for 
useful comments that helped improve the presentation.


\end{document}